\documentclass[twocolumn,superscriptaddress,showpacs]{revtex4}  
\usepackage{ae}
\usepackage[T1]{fontenc}
\usepackage[ansinew]{inputenc}
\usepackage{amsmath}
\usepackage{graphicx}
\usepackage{color}
\usepackage[colorlinks]{hyperref}
\usepackage{setspace}

\begin{document}

\date{\today}
\title{Nanoelectromechanics of shuttle devices}
\author{R. I. Shekhter}
\affiliation{Department of Physics, University of Gothenburg, SE-412
96 G{\" o}teborg, Sweden}
\author{L. Y. Gorelik}
\affiliation{Department of Applied Physics, Chalmers University of
Technology, SE-412 96 G{\" o}teborg, Sweden}
\author{I. V. Krive}
\affiliation{B. Verkin Institute for Low Temperature Physics and
Engineering of the National Academy of Sciences of Ukraine, 47 Lenin
Ave., Kharkov 61103, Ukraine} \affiliation{Physical Department, V.
N. Karazin National University, Kharkov 61077, Ukraine}
\author{M. N. Kiselev}
\affiliation{The Abdus Salam International Centre for Theoretical
Physics, Strada Costiera 11, 1-34151 Trieste, Italy}
\author{A. V. Parafilo}
\affiliation{B. Verkin Institute for Low Temperature Physics and
Engineering of the National Academy of Sciences of Ukraine, 47 Lenin
Ave., Kharkov 61103, Ukraine}
\author{M. Jonson}
\affiliation{Department of Physics, University of Gothenburg, SE-412 96 G{\" o}teborg, Sweden}
\affiliation{SUPA, Institute of Photonics and Quantum Sciences, Heriot-Watt University, Edinburgh EH14 4AS, Scotland, UK}
\affiliation{Department of Physics, Division of Quantum Phases and
Devices, Konkuk University, Seoul 143-701, Korea}

\date{\today}
\pacs{73.23.-b,
  72.10.Fk,
  73.23.Hk,
  85.85.+j}

\begin{abstract}
A single-electron tunneling (SET) device with a nanoscale central
island that can move with respect to the bulk source- and drain
electrodes allows for a nanoelectromechanical (NEM) coupling between
the electrical current through the device and mechanical vibrations
of the island. Although an electromechanical "shuttle" instability
and the associated phenomenon of single-electron shuttling were
predicted more than 15 years ago, both theoretical and experimental
studies of NEM-SET structures are still carried out. New
functionalities based on quantum coherence, Coulomb correlations and
coherent electron-spin dynamics are of particular current interest.
In this article we present a short review of recent activities in
this area.
\end{abstract}
\maketitle

\section{Introduction} \label{int}

Electro-mechanical and mechano-electrical transductions phenomena
have historically contributed greatly to the advancement of
technology in our society. Today, such operations can be achieved on
the single-molecular level with obvious advantages brought about by
the miniaturization of the devices involved. In addition
qualitatively novel functionalities become available due to new
physics that becomes relevant in materials structured on the
nanometer length scale. Quantum mechanics and electron-electron
(Coulomb) correlations are defining ingredients of mesoscopic
physics, which applies to nanoscale devices whose properties may be
determined by a single or a few degrees of freedom. As a result
there is a possibility and also an advantage to be gained by taking
quantum coherence into account when designing electro-mechanical
devices for the purpose of quantum manipulation and quantum
communication.

The single-electron tunneling (SET) transistor is a nanodevice with
particularly prominent mesoscopic features.
Here, the Coulomb blockade of single-electron tunneling at low
voltage bias and temperature \cite{coulombblockade} makes Ohm's law for the
electrical conductance invalid in the sense that the electrical
current is not necessarily proportional to the voltage drop across
the device.
Instead, the current  is due to a temporally discrete set of events where
electrons tunnel quantum-mechanically one-by-one from a source to a
drain electrode via a nanometer size island (a ``quantum dot").
This is why the properties of a single electronic quantum state are
crucial for the operation of the entire device.

Since the probability for quantum mechanical tunneling is
exponentially sensitive to the tunneling distance it follows that
the position of the quantum dot relative to the electrodes is
crucial. On the other hand the strong Coulomb forces that accompany
the discrete nanoscale charge fluctuations, which are a necessary
consequence of a current flow through the SET device, might cause a
significant deformation of the device and move the dot, hence giving
rise to a strong electro-mechanical coupling. This unique feature
makes the so-called nanoelectromechanical SET (NEM-SET) devices, where
mechanical deformation can be achieved along with electronic
operations,
to be one of the best nanoscale realizations of
electromechanical transduction.

In this review we will discuss some of the latest achievements in the
nano-electromechanics of NEM-SET devices focusing on the new
functionality that exploits quantum coherence in both the electronic
and the mechanical subsystems. The choice of materials for making a
NEM-SET device brings an additional dimension to exploring its
quantum performance. By choosing superconductors or magnets as
components of the device one may, e.g., take advantage of a
macroscopic ordering of electrons with respect to both their charge
and spin. We will discuss how the electronic charge as well as the
electronic spin contribute to electromechanical and
mechano-electrical transduction in a NEM-SET device. New effects
appear also due to the high mechanical deformability of molecular
NEM-SET structures (polaronic effects) which could be accompanied
with effects of strong electron-electron interactions (Coulomb
blockade and Luttinger liquid phenomena) as well as effects caused
by strong tunneling coupling (Kondo- nano-mechanics).

The phenomenon of shuttling and the sensitivity of electronic
tunneling probabilities to mechanical deformation of the device are
in the focus of the present review. In this sense it is is an update
of our earlier reviews of shuttling \cite{shuttlerev1, shuttlerev2,
shuttlerevfnt}. Other aspects of nanoelectromechanics are only
briefly discussed here. We refer readers to the well-known reviews
of Refs.~\onlinecite {blencowe, roukes1, roukes2,cleland,zant} on
nanoelectromechanical systems for additional information.

This review is organized as follows. In Section II we introduce the basic
concepts of electron shuttling (Subsection II.A) and consider the
influence of polaronic effects on the shuttle instability (Subsection
II.B). In the end of this subsection the quantum shuttle is introduced and
briefly discussed. Section III deals with the shuttling of Cooper
pairs (Subsection III.A) and polaronic effects on the Josephson current
through a vibrating quantum dot (Subsection III.B). A novel phenomenon ---
magnetic shuttling --- is considered in Section IV, where the effects of
spin-controlled shuttling of electric charge (Subsection IV.A), the
spintro-mechanics of the magnetic shuttle (Subsection IV.B) and mechanical
transportation of magnetization (Subsection IV.C) are discussed. In the
end of Section III the Kondo regime of electron shuttling is
reviewed (Subsection IV.D). Recent experiments on the observation of
electron shuttling are briefly discussed in Section V. Section VI
summarizes the latest theoretical achievements in
nanoelectromechanics of shuttle devices.

\section{Shuttling of single electrons}

A single-electron shuttle can be considered as the ultimate
miniaturization of a classical electric pendulum capable of
transferring macroscopic amounts of charge between two metal plates.
In both cases the electric force acting on a charged ``ball" that is
free to move in a potential well between two metal  electrodes kept
at different electrochemical potentials, $eV=\mu_L-\mu_R$, results
in self-oscillations of the ball. Two distinct physical phenomena,
namely the quantum mechanical tunneling mechanism for charge loading
(unloading) of the ball (in this case more properly referred to as a
grain) and the Coulomb blockade of tunneling, distinguish the
nanoelectromechanical device known as a single-electron shuttle
\cite{shuttleprl} (see also \cite{electrinst}) from its classical
textbook analog. The regime of Coulomb blockade realized at bias
voltages and temperatures $eV, T \ll E_C$ (where $E_C= e^2/2C$ is
the charging energy, $C$ is the grain's electrical capacitance)
allows one to consider single electron transport through the grain.
Electron tunneling, being extremely sensitive to the position of the
grain relative to the bulk electrodes, leads to a shuttle
instability --- the absence of any equilibrium position of an
initially neutral grain in the gap between the electrodes.

In Subsection A below we consider the characteristic features of the
single-electron shuttle in the case when the shuttle dynamics can be treated as
classical motion. Quantum corrections to this picture are then discussed in
Subsection B.

\subsection{Shuttle instability in the quantum regime of Coulomb blockade}

Theoretically, it is convenient to  study the single-electron
shuttle in an approach \cite{shuttlefedor} where the grain is
modeled as a single-level  quantum dot (QD) that is weakly  coupled
(via a tunnel Hamiltonian) to the electrodes (see Fig. 1). The
Hamiltonian corresponding to this model reads
\begin{figure}
\vspace{0.cm} {\includegraphics[width=8cm]{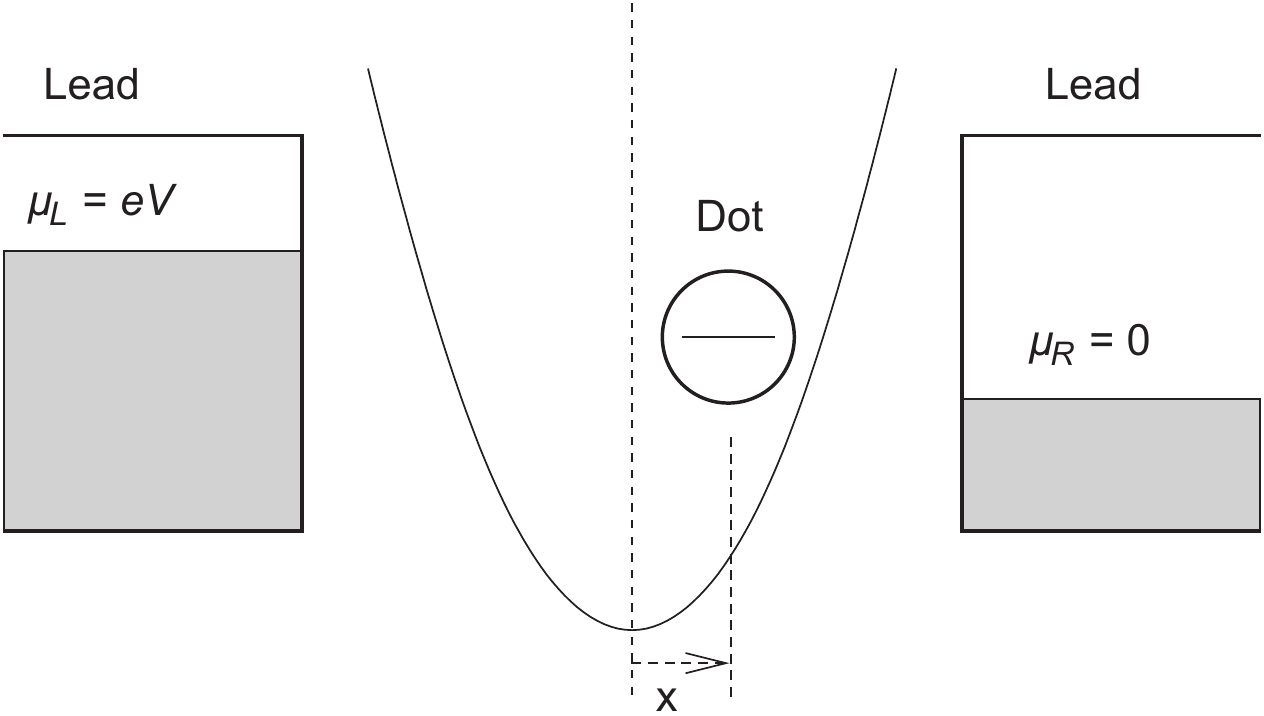}} \vspace*{0.5
cm} \caption{Model system consisting of a movable quantum dot placed
between two leads. An effective elastic force acting on the dot due
to its connections to the leads is described by a parabolic
potential. Only one single electron state is available in the dot
and the non-interacting electrons in the leads are assumed to have a
constant density of states. Reprinted with permission from
\cite{shuttlefedor}, D. Fedorets {\em et al.},  { Europhys. Lett.}
{\bf 58}, 99 (2002). $\copyright$ 2002, EDP Sciences. }
\end{figure}
\begin{equation} \label{Model}
H=\sum_{j=L,R}H^{(j)}_l+H_{QD}+\sum_{j=L,R}H_t^{(j)},
\end{equation}
where the Hamiltonian
\begin{equation} \label{lead}
H_l^{(j)}=\sum_k(\varepsilon_{kj}-\mu_j)a^{\dag}_{kj}a_{kj}
\end{equation}
describes noninteracting electrons in the left ($j=L$) and right
($j=R$) leads, which are kept at different chemical potential
$\mu_j$; $a^{\dag}_{kj}(a_{kj})$ creates (annihilates) an electron
with momentum $k$ in  lead $j$. The QD Hamiltonian takes the form
\begin{equation} \label{QD}
H_{QD}=\varepsilon_0d^{\dag}d+\varepsilon_i\hat{X}d^{\dag}d+\frac{\hbar\omega_0}{2}(\hat{X}^2+\hat{P}^2),
\end{equation}
where $d^{\dag}(d)$ is the creation (destruction) operator for an
electron on the dot, $\varepsilon_0$ is the energy of the resonant
level, $\hat{X}$ is the dimensionless coordinate operator
(normalized by the amplitude $x_0$ of zero-point fluctuations,
$x_0=\sqrt{\hbar/M\omega_0}$, $M$ is the mass of QD), $\hat{P}$ is
the corresponding momentum operator ($[\hat{X},\hat{P}]=i$), $\omega_0$ is
the frequency of vibrons and $\varepsilon_i$ is the
electromechanical interaction energy. The physical meaning of the
second term in Eq.~(\ref{QD}) is the interaction energy due to the
coupling of the electron charge density on the dot with the electric
potential $\phi(x)=E x_{cm}$, where $E=V/d_0$ is the electric field
in the gap between electrodes ($d_0$ is the distance between the
electrodes, $x_{cm}$ is the center-of-mass coordinate of the quantum
dot). In this case the coupling "constant" $\varepsilon_i$ from
Eq.~(\ref{QD}) is a linear function of bias voltage
$\varepsilon_i(V)\simeq e V x_0/d_0$.

The tunneling Hamiltonian $H_t^{(j)}$ in Eq.~(\ref{Model}) differs
from its standard form. The explicit coordinate dependence of the
tunneling matrix elements introduces additional electron-vibron interactions
(additional to those described by the second term of Eq.~(\ref{QD}). These
result in the appearance of quantum cohesive forces
$\hat{F}_c^{(j)}=-\partial\hat{H}_t^{(j)}/\partial\hat{X}$,
$j=(+,-)\equiv(L,R)$, where
\begin{equation} \label{tunnel}
H_t^{(j)}=\sum_kt_{0j}\exp(j\lambda_t\hat{X})a_{kj}^{\dag}d+h.c.
\end{equation}
Here $t_{0j}$ is the bare tunneling amplitude, which corresponds to
a weak dot-electrode coupling, $\lambda_t=x_0/l_t$ is a
dimensionless parameter ($l_t$ is the electron tunneling length)
that characterizes the sensitivity of the tunneling matrix elements
to a shift of the dot center-of-mass coordinate with respect to its
equilibrium ($x_{cm}=0$) position.

In this Section we are interested in the classical shuttle motion
$\langle\hat{X}\rangle=x_c(t)\gg1$ induced by an applied voltage
$V$. Here the average
$\langle{\cal O}\rangle\equiv\textrm{Tr}\{\hat{\rho}(t){\cal O}\}$
is taken with the statistical operator $\hat{\rho}(t)$ obeying the
Liouville-von Neumann equation
\begin{equation} \label{LvN eq}
i\hbar\partial_t\hat{\rho}(t)=[\hat{H},\hat{\rho}(t)].
\end{equation}

For a weak electron-vibron interaction characterized by the
dimensionless coupling constant
$\lambda=-\sqrt{2}\varepsilon_i/\hbar\omega_0\ll1$, one can neglect
the effects ($\sim\lambda^2$) of zero-point QD fluctuations (see
below). In this case the classical dot coordinate $x_c(t)$ is
governed by Newton's equation \cite{shuttlefedor}
\begin{equation} \label{Newton}
\ddot{x}_c+\omega_0^2x_c=F(t)/M,
\end{equation}
where the average force
$F(t)=-\textrm{Tr}\{\hat{\rho}(t)(\partial\hat{H}/\partial\hat{X})\}=F_e(t)+F_c(t)$
 consists of two terms: the electric force $F_e\propto\lambda$ acting
 on the accumulated charge on the QD, and the cohesive force,
 $F_c\propto\lambda_t$, produced by the position-dependent
 hybridization of the electronic states of the grain and the leads.
 The evaluation of both forces can be done analytically
 \cite{shuttlefedor, fedorets},
 either by solving Heisenberg equations of motion for the fermion operators ($a_{kj}$, $d$) or
 by using the Keldysh Green's function approach. The nonlinear
 Eq.~(\ref{Newton}) for the classical shuttle motion  can be analyzed in two
 cases: (i) near the shuttle instability ($x_c\rightarrow0$), and (ii) for the developed
 shuttle motion (finite $x_c$, small $\lambda_tx_c$).

 For weak electromechanical coupling $\lambda,\lambda_t\ll1$ it was
 shown \cite{shuttlefedor} that the amplitude of initially small oscillations starts to grow exponentially ($\sim e^{r_st}$)
 if $eV>eV_c=2(\varepsilon_0
 +\hbar\omega_0)$, which means that the threshold voltage for the shuttle instability is $V_c$.
 \begin{figure}
\vspace{0.cm} {\includegraphics[width=8cm]{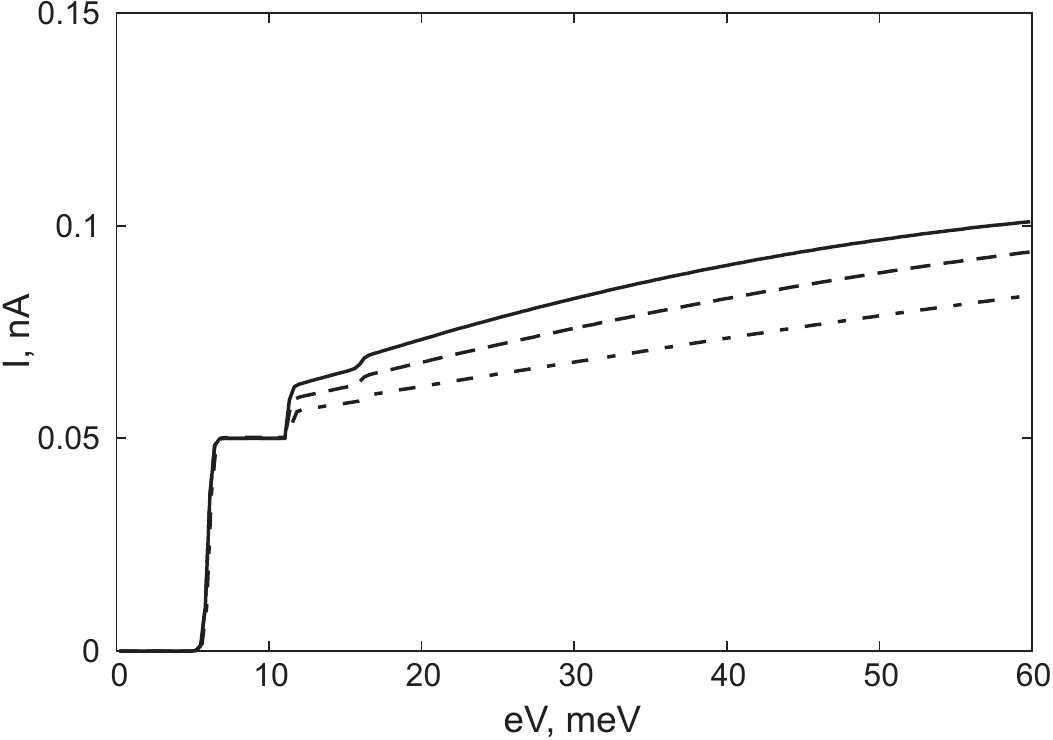}} \vspace*{0.
cm} \caption{Step-like I-V curves for single-electron shuttling for
different model parameters. Reprinted with permission from
\cite{shuttlefedor}, D. Fedorets {\em et al.},  { Europhys. Lett.}
{\bf 58}, 99 (2002). $\copyright$ 2002, EDP Sciences.}
\end{figure}
The rate of growth of the instability, $r_s$, depends on the level
width $\Gamma=\Gamma_L\Gamma_R/(\Gamma_L+\Gamma_R)$, where
$\Gamma_j=2\pi\nu_0|t_{0j}|^2$ ($\nu_0$ is the density of states, which
is assumed to be an energy independent quantity in the wide band
approximation \cite{glazshekhter, wba}) and on the strengths of the
electromechanical coupling $r_s\sim\lambda\lambda_t\Gamma/\hbar$. It
was also shown  \cite{shuttlefedor} that even in the absence  of
mechanical friction, which can be taken into account
phenomenologically by adding to Eq.~(\ref{Newton}) the dissipative
term $\gamma\dot{x}_c$, the instability develops into a limit cycle.
This is in contrast with a classical shuttle \cite{shuttleprl},
where stability of the system can be achieved, only at finite
mechanical dissipation $\gamma\neq0$. In the considered model the
effective dissipation is provided by the hybridization
($\Gamma_{L(R)}$) of the resonant level with the metallic leads.

Notice that the cohesive force $F_c$ is unimportant for the
developed shuttle motion. However, this exchange interaction along
with the direct electric coupling ($\lambda$) determines the growth
rate of the shuttle instability, which demonstrates the important
role of electron tunneling in the dynamics of single electron
shuttling.

\subsection{Strong electron-vibron interaction and polaronic effects in electron shuttling}

What is the role of quantum effects in the vibrational subsystem? It
is known that the electron-vibron interaction (second term in
Eq.~(\ref{QD})) results in vibron-assisted electron tunneling
\cite{glazshekhter} (the appearance of inelastic channels) and
for $\lambda\gg1$ in strong suppression  of the probability of electron
transfer through the elastic channel (Franck-Condon blockade
\cite{116}, see also the reviews \onlinecite{ratner, palevski}). Could
these vibrational effects influence the single-electron shuttling
phenomenon?

\begin{figure}
\vspace{0.cm} {\includegraphics[width=8cm]{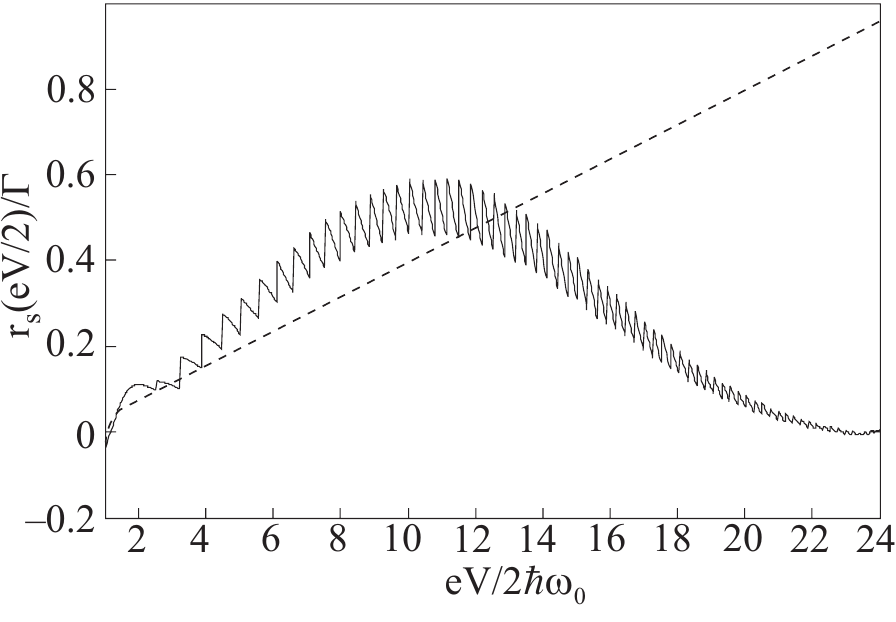}} \vspace*{-0.
cm} \caption{The increment rate of the shuttle instability as a
function of bias voltage for $T/\hbar\omega_0=0.2$. The dotted line
represents the result of Ref.~\onlinecite{fedorets} extended to the
region of strong electromechanical coupling. Reprinted with
permission from \cite{gleb}, G.A. Skorobagat'ko {\em et al.},  Fiz.
Nizk. Temp. {\bf 35}, 1221 (2009). $\copyright$ 2009, B. Verkin
Institute for Low Temperature and Engineering of the NAS of Ukraine.
}
\end{figure}

The problem just stated was considered in Ref.~\onlinecite{gleb}. Quantum
fluctuations of the dot coordinate can be taken into account by
replacing the operator $\hat{X}$ in Eqs.~(\ref{QD},\ref{tunnel}) by
$x_c(t)+\hat{x}$, where $\langle\hat{x}\rangle=0$. The ``quantum"
part of the electron-vibron interaction
($\varepsilon_i\hat{n}\hat{x}$) in Eq.~(\ref{QD}) can be eliminated
in the Hamiltonian (\ref{tunnel}) by a standard trick - the
Lang-Firsov unitary transformation \cite{mahan, langfirsov}
$U=\exp(i\lambda\hat{p}\hat{n})$, where $[\hat{x},\hat{p}]=i$. After
the unitary transformation the electron-vibron coupling appears in
the tunneling Hamiltonian in the form of an additional operator
factor $\exp(i\lambda\hat{p})$. Since in the transformed Hamiltonian
electron-vibron interactions enter only in the tunneling Hamiltonian,
the problem can be solved in perturbation theory with respect to the bare tunneling
width $\Gamma_0$. In the lowest order of perturbation theory the
averages of bosonic and fermionic operators are decoupled and the
bosonic correlation functions
$\langle\exp[\alpha^{\ast}b^{\dag}(t)+\beta^{\ast}b(t)]\exp[\alpha
b(0)+\beta b^{\dag}(0)]\rangle_0$ can be evaluated analytically
($\langle...\rangle_0$ denotes an average with respect to the
noninteracting Hamiltonian $H_0=\hbar\omega_0b^{\dag}b$). The result
\cite{gleb} is an increment rate $r_s$ of the shuttle instability that is
valid both for weak $\lambda\ll1$ and strong
$\lambda\gtrsim1$ electromechanical coupling. For a symmetric
junction $\Gamma_L=\Gamma_R=\Gamma_0$ and $T=0$ it reads
\begin{equation} \label{increment}
r_s\simeq\frac{\Gamma_0}{\hbar}\lambda\lambda_t\exp(-\lambda^2+\lambda_t^2-\lambda\lambda_t)\sum_{l=0}^{l_m-1}\frac{(\lambda+\lambda_t
)^2}{l!},
\end{equation}
where
\begin{equation} \label{where}
l_m=\left[\frac{eV}{2\hbar\omega_0}-\left(\frac{\varepsilon_0}{\hbar\omega_0}-\frac{\lambda^2}{2}\right)-1\right]
\end{equation}
and the symbol $[\bullet]$ here denotes the integer part. In the
limit of weak electromechanical couplings $\lambda\ll1$,
$\lambda_t\ll1$ Eq.~(\ref{increment}) reproduces the result of
Ref.~\onlinecite{shuttlefedor},
$r_s\approx\Gamma_0\lambda\lambda_t/\hbar$, which predicts a linear
dependence of $r_s$ on the bias voltage, $r_s\propto\lambda\propto
V$. Two new factors in Eq.~(\ref{increment}) (the exponential factor
and the sum over open inelastic channels) reflect the influence of
two major vibrational effects - the Franck-Condon blockade and
vibron-assisted electron tunneling - on the shuttle instability.
Since the dimensionless coupling $\lambda_t=x_0/l_t$ is always small
for molecular devices, the main quantum effect of vibrations is the
``polaronic" narrowing of the bare tunneling width
$\Gamma_0\rightarrow\Gamma_0\exp(-\lambda^2(V))$. It increases with
the increase of the number of new channels ($eV\gg\hbar\omega_0$)
and reaches its maximum at $eV_m\sim M (d\omega_0)^2$ ($d$ is the
distance between electrodes, $M$ is the mass of the vibrating
molecule). A nonmonotonic dependence of the increment rate of the
shuttle instability on $V$ is shown in Fig.~3. One can see that an
instability takes place in a finite interval of bias voltages due to
the Franck-Condon blockade. Besides, at low temperatures,
$T\ll\hbar\omega_0$, the increment rate oscillates with a period of
order $\hbar\omega_0$ and with a relatively large amplitude. Notice
that at low voltages $\lambda(V)\ll1$ and hence ``polaronic" effects
could not affect the ``intrinsic" shuttle instability, which takes
place at $V=2\hbar\omega_0/e$. However, if the grain is pinned or if
there is strong friction in the system, $\gamma\gg\omega_0$, the
shuttle instability takes place at much higher voltages when
$\lambda(V)\gtrsim1$. In this case large oscillations of the
increment parameter could lead to unusual behavior of the I-V
characteristics for a shuttle-based single-electron transistor. A
small change in bias voltage (smaller than $\hbar\omega_0/e$) would
take the system from the shuttle regime of transport (with strongly
enhanced electron tunneling probability) to the ordinary regime of
tunnel transport (small tunneling probability) and the other way
around. Therefore one can expect pronounced negative differential
conductance (NDC) features (on the scale of $\hbar\omega_0/e$) in
the current voltage characteristics (see also Refs.~\onlinecite{111,
116}).

It is interesting to note that in the absence of mechanical
damping, ($\gamma=0$), the threshold voltage for single electron
shuttling is determined by the vibron energy \cite{shuttlefedor}
$eV_{th}=2\hbar\omega_0$
 This quantum threshold value
indicates where the process of inelastic electron tunneling (with
emission of a vibron) becomes energetically available. In the
realistic case of finite friction the threshold bias voltage is
found by solving the equation $r_s(V_{th})=\gamma$, where $r_s(V)$
is defined in Eq.~(\ref{increment}) (see Fig. 3). In the limit of
weak electromechanical coupling one gets for the threshold electric
field ($E_{th}=V_{th}/d$) the result
\begin{equation} \label{threshold}
eE_{th}=\frac{\hbar\gamma}{\Gamma}M \omega_0l_t,
\end{equation}
which has a linear dependence on the rate of dissipation $\gamma$.
The exponential increase (above threshold) of the amplitude $A(t)$
of the shuttle motion means that for the fully developed
(stationary) shuttle motion $A(\infty)\gg x_0$. However, at the
initial stage of the instability the oscillation amplitude can be of
the order of $x_0$ and the classical treatment of shuttle motion
ceases to be valid.

A fully quantum-mechanical approach to single electron shuttling was
developed in Refs.~\onlinecite{fundam, novotny, quanshuttle}, where it was
shown that the shuttle instability (exponential increase of
$\langle\hat{x}\rangle$ and $\langle\hat{p}\rangle$) in the limit
$l_t\gg l_E=eE/M\omega_0^2\sim x_0$ occurs for a threshold electric
field that coincides with  Eq.~(\ref{threshold}), found in a
quasiclassical approach. By using the Wigner distribution function,
which allows one to visualize the behavior of a quantum system in
phase space \cite{novotny}, two different regimes of single electron
shuttling were found. The classical regime (small fluctuations
around the stationary trajectory $x_c(t)=A_c\sin(\omega_0 t)$) is
realized for the fields $E\gg E_q>E_{th}$, where \cite{quanshuttle}
\begin{equation} \label{field}
eE_q=C\left(\frac{x_0}{l_t}\right)^4M\omega_0^2l_t
\end{equation}
($C\simeq10^{-2}$), i.e. in the case of weak mechanical dissipation
$\gamma\lesssim\Gamma(x_0/l_t)^4$ and for large bias voltages. For
low biases, when the electric fields, acting on the charged QD are
in the interval $E_{th}<E<E_q$, the shuttle regime has a specific
quantum character. The Wigner function is strongly smeared around
the classical trajectory. It is concentrated in a region between two
circles with radii $R_{out}, R_{in}\gg x_0$ and
$R_{out}-R_{in}\gg x_0$ for $l_t\gg x_0$. This behavior is
characterized by pronounced quantum fluctuations and can be
interpreted as a quantum shuttle. It is difficult to detect a
quantum shuttle  by measuring the average current since its
qualitative behavior has no distinctive features in comparison with
the classical shuttle. The noise properties (and in general the full
counting statistics) of NEMS will be crucial for detecting single-electron shuttling.

\section{Mechanically mediated superconductivity and polaronic effects in the Josephson
current} \label{sup}

Shuttling of electric charges between nonsuperconducting electrodes
by itself does not require phase coherence. Even in the quantum
regime of the Coulomb blockade, when only a single (resonant) level is
involved in electron transport, phase coherent effects have little
influence on electron shuttling. This is not the case for magnetic
and superconducting leads. Magnetic exchange forces make the coherent
electron-spin dynamics important for electron shuttling.
Superconducting transport is by definition a phase coherent
phenomenon and thus Cooper-pair shuttling has to be strongly
different from single-electron shuttling.

In this Section we consider shuttling of Cooper pairs between two
superconducting electrodes (subsection A) and the influence of
vibrational modes on the Josephson current.

\subsection{Shuttling of Cooper pairs}
 The main requirement for the observation of a
mechanically mediated Josephson current \cite{2.5, 2.6} is that
phase coherence is preserved during the transportation of Cooper
pairs between the two superconducting leads, and during the process
of transferring charge between the bulk superconductors and a
movable superconducting grain. This requirement can be fulfilled if
the superconducting grain is small enough to be in the Coulomb
blockade regime \cite{coulombblockade} (see also \cite{vanhouten})
so that it can play the role of a single-Cooper-pair box \cite{2.2}
(see also \cite{naka, devo}). The implication is that the
characteristic energy scales of the small superconducting grain --
the Josephson energy $E_J=(\hbar^2/2 e)J_c$ (where $J_c$ is the
critical current) and the charging energy $E_C=(2e)^2/2C$ (where $C$
is the grain capacity) -- have to obey the double inequality $E_J\ll
E_C \ll \Delta$ (where $\Delta$ is the superconducting gap) while
the temperature has to be low enough to make $T\ll E_C$. In this
regime the single-electron states on the grain are energetically
unfavorable (the parity effect \cite{quant,2.2}) and the
superconducting properties of the system can be described by a
two-level model (see for example the review \cite{2.1}). The
corresponding state vector of a single-Cooper-pair box is a coherent
superposition of the states with $n=0$ and $n=1$ Cooper pairs ({
more generally states with the different number of Cooper pairs:
$2N$ and $2(N+1)$}) on the grain. For a movable Cooper pair box the
energy scale for mechanical vibrations $\hbar\omega_0$ has to be
much smaller then all other energy scales. This additional
requirement prevents the creation of quasiparticles and allows one
to consider the mechanical motion of the grain as an adiabatic
process.

{ The Hamiltonian of the system is expressed in terms of the Cooper
pair number operator $\hat{n}$ for the grain and the phases of the
superconducting leads, $\varphi_{L,R}$ }:
\begin{equation} \label{CooppairHam}
H=-\frac{1}{2}\sum_{i=L,
R}E^i_J\{x(t)\}[e^{i\varphi_{L,R}}|1\rangle\langle0|+h.c.]+\delta
E_C\{x(t)\}\hat{n}.
\end{equation}
{ The operator $|1\rangle\langle0|$ changes the number of Cooper
pairs on the grain. An essential specific feature here is the
dependence of the charging energy difference $\delta
E_C=E_C(n=1)-E_C(n=0)$ and the coupling energies
$E^{L,R}_{J}(x)=E_0\exp(-\delta x_{L,R}/l_t)$ ($\delta x_{L,R}$ is
the distance between the grain and the respective lead) on the
instantaneous position $x(t)$ of the superconducting grain.}

It is useful to separate the adiabatic motion of the
single-Cooper-pair box between the two superconducting electrodes
into two different parts: (i) the free motion ({ transportation
region in Fig. 4}), and (ii) the process of loading and unloading of
charge 
near the leads ({ contact region in Fig. 4}). During the free
motion, {when the Josephson energy is negligibly small and the
Coulomb term dominates}, the dynamics of the qubit is reduced to the
time evolution of the relative phase $\chi$ { due to the second term
in Eq. (\ref{CooppairHam})}, $\hbar\dot{\chi}=\delta E_C$. In
general the accumulated phases are different for left-to-right
($t_{+}$) and right-to-left ($t_{-}$) motion; $\chi_{\pm}\simeq
\delta E_C t_{\pm}/\hbar$. The coherent exchange of a Cooper pair
between the grain and the lead [stage (ii)] is characterized by the
dimensionless Josephson coupling strength $\theta_J\simeq
E_Jt_c/\hbar$, where $t_c$ is the time spent by the grain in contact
with the lead (see Fig. 4). The superconducting phase difference
$\varphi=\varphi_R-\varphi_L$, the dynamical phases $\chi_{\pm}$ and
the Josephson coupling strength $\theta_J$ fully control the
behavior of the Josephson current. The characteristic value of the
mechanically assisted supercurrent (the ``critical" current $J_m$)
for a periodic  motion of the grain (with frequency
$f=\omega_0/2\pi$) and for strong Josephson coupling
$\theta_J\approx1$ is determined by the mechanical frequency only,
$J_m\simeq 2 e f$. An analytical expression for the mechanically
mediated dc Josephson current was derived in Ref.~\onlinecite{2.5}
and takes the form
\begin{equation} \label{dccur}
J=2 e f \frac{\sin^3\theta_J \cos\theta_J \sin\Phi
(\cos\Phi+\cos\chi)}{1-(\cos^2\theta_J \cos\chi-\sin^2\theta_J
\cos\Phi)^2},
\end{equation}
where $\Phi=\varphi+\chi_+-\chi_-$, $\chi=\chi_++\chi_-$. The
current Eq.~(\ref{dccur}) is an oscillating function of the
superconducting phase difference $\varphi$ (see Fig. 5), which is a
spectacular manifestation of a Josephson coupling between the remote
superconductors. In the limit of weak coupling $\theta_J\ll 1$ and
vanishingly small dynamical phase ($\chi\rightarrow0$)
Eq.~(\ref{dccur}) is reduced to the standard Josephson formula
$J=J_c\sin\varphi$, where $J_c\simeq e E_J/\hbar$. We see from
Eq.~(\ref{dccur}) that the main qualitative effect of the dynamical
phase, which can be controlled by the gate voltage, is a change of
the direction of supercurrent (if $\cos\chi+\cos\Phi<0$). For a
given strength of the Josephson coupling the direction of a mechanically
mediated supercurrent is determined by the interplay of
superconducting ($\varphi$) and dynamical ($\chi$) phases. Notice
that in Ref.~\onlinecite{2.6} it was shown that mechanical
transportation of Cooper pairs could establish relative phase coherence
between two mesoscopic superconductors if initially they are in states
with strong uncorrelated phase fluctuations.

\begin{figure}
\vspace{0.cm} {\includegraphics[width=8cm]{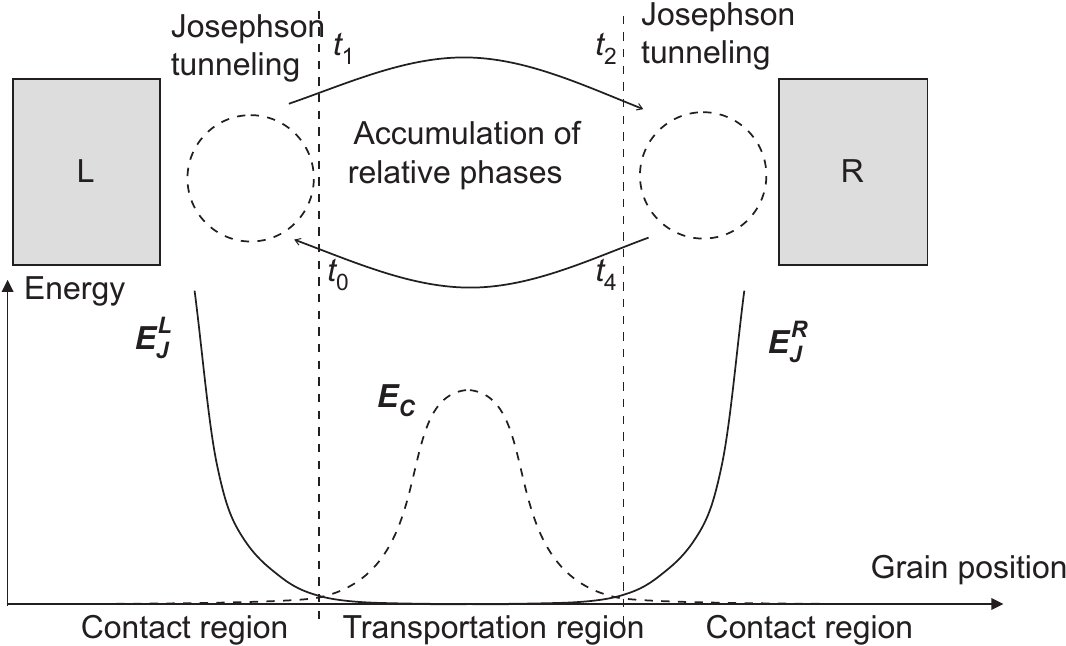} } \vspace*{-0.
cm} \caption{Illustration of the charge transport process. The
central island moves periodically between the leads.
Close to each turning point Cooper pair tunneling between lead and
island is possible since the voltage on a gate electrode (not shown)
has been set to locally remove the electrostatic energy difference
between having zero or one extra Cooper pair on the island (i.e.,
the difference in charging energy, $\delta E_C$, is zero
\cite{footnote}). As the island retracts from the lead, tunneling is
exponentially suppressed ($E_J=0$) while the degeneracy of the two
charge states is lifted as the influence of the gate is weakened
($\delta E_C\ne 0$) \cite{2.5}. Reprinted with permission from
\cite{shuttlerev1}, R. I. Shekhter {\em et al.},  { J. Phys.:
Condens. Matter} {\bf 15}, R441 (2003). $\copyright$ 2003, Institute
of Physics and IOP Publishing Limited. }
\end{figure}

\begin{figure}
\vspace{0.cm} {\includegraphics[width=7cm]{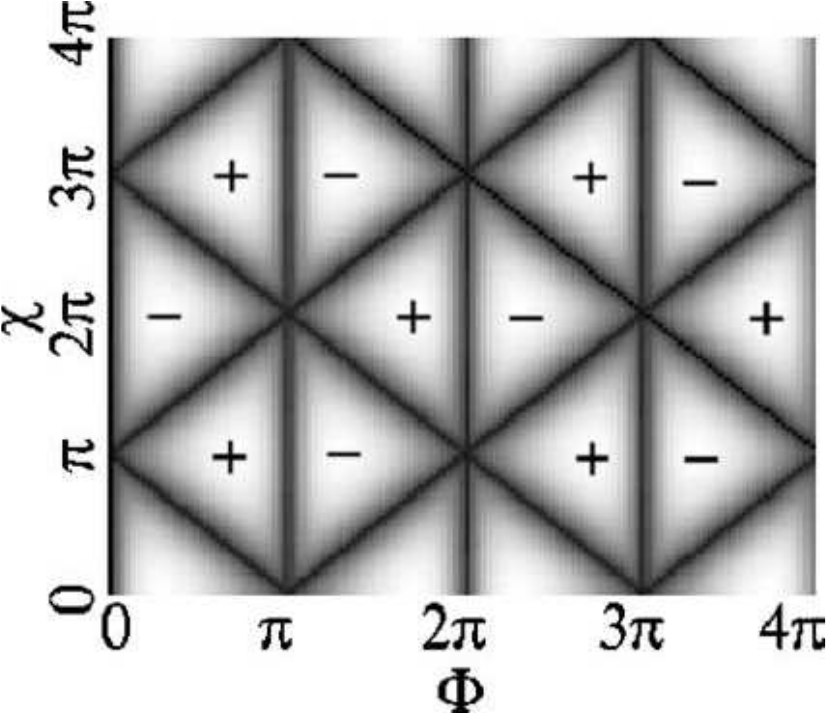}} \vspace*{-0.
cm} \caption{Magnitude of the current in Eq.~(\ref{dccur}) in units
of $2ef$ as a function of the phases $\Phi$ and $\chi$. Regions of
black correspond to no current and regions of white to $J/2ef=0.5$.
The direction of the current, is indicated in by signs ($\pm$). To
best see the ``triangular" structure of the current, the Josephson
coupling has been chosen to be $\theta_J=\pi/3$. Reprinted with
permission from \cite{shuttlerev1}, R. I. Shekhter et al.,  { J.
Phys.: Condens. Matter} {\bf 15}, R441 (2003). $\copyright$ 2003,
Institute of Physics and IOP Publishing Limited.}
\end{figure}

\subsection{Josephson current through a vibrating quantum dot}

In Section 2 we considered the influence of polaronic effects on the
electron shuttle instability. Here we analyze how vibrational
degrees of freedom affect the dc Josephson current. When studying
the transport properties of a superconductor/quantum
dot/superconductor (SQDS) junction we will use the same simple model
for the QD as in the previous Section, i.e. we consider a dot with a single
energy level $\varepsilon_0$ that vibrates with angular frequency
$\omega_0$ and is weakly coupled to the superconducting leads. The
latter are described by the standard BCS Hamiltonian with order
parameter $\Delta_j=\Delta_0e^{i\varphi_j}$ ($\Delta_0$ is the
superconducting gap and $\varphi_j$ is the phase of the order parameter
for the left, $j=L$, and right, $j=R$, superconductor). The coupling
of the dot to the leads is described by a tunnel Hamiltonian,
which introduces two energy scales to the problem, viz. the partial level widths
$\Gamma_{L,R}$. These are very significant for the transport properties.

The QD Hamiltonian reads
\begin{equation} \label{QD HAm}
H_{QD}=\sum_{\sigma}\varepsilon_0d^{\dag}_{\sigma}d_{\sigma}+\varepsilon_i\hat{n}(b+b^{\dag})/\sqrt{2}+
U_C\hat{n}_{\uparrow}\hat{n}_{\downarrow}+\hbar\omega_0b^{\dag}b,
\end{equation}
where $d_{\sigma}(d^{\dag}_{\sigma})$ is the destruction (creation)
operator for an electron with spin projection
$\sigma=\uparrow,\downarrow$,
$\hat{n}=\hat{n}_{\uparrow}+\hat{n}_{\downarrow}$,
$\hat{n}_{\sigma}=d^{\dag}_{\sigma}d_{\sigma}$, $b(b^{\dag})$ is the
vibron destruction (creation) operator, $\varepsilon_i$ is the
 electron-vibron interaction energy and $U_C$ is the
electron-electron interaction energy. In what follows we will assume
that $\Delta_0$ is the largest energy scale in the problem. This
allows one to neglect quasiparticles (continuum spectrum) when
calculating the dc Josephson current.

In general, a coupling to vibrational modes tends to suppress the
supercurrent \cite{flensberg,shumeikospec}. The suppression
mechanism
 is different for ``hard", $\hbar\omega_0\gg \Gamma$, and
``soft", $\hbar\omega_0\ll \Gamma$, vibrons. {For hard vibrons and
for $\hbar\omega_0\gtrsim\Delta_0$ only the ground state of the
vibrational subsystem is involved in Cooper pair transport through a
S/QD/S junction}. Zero-point fluctuations of the QD result in strong
(exponential) renormalization of the electron tunneling probability
(``polaronic" narrowing of the level width). For strong
electron-vibron coupling
--- i.e. for $\lambda\gtrsim 1$,
 the critical current is exponentially
suppressed \cite{flensberg}, which is a manifestation of the
Franck-Condon blockade \cite{ratner, 116} of the supercurrent. {
Notice, that the Franck-Condon blockade will be partially removed
when $\hbar\omega_0\ll\Delta_0$ due to contributions of virtual
side-band channels}.

The effect { of the Franck-Condon blockade } was first predicted
\cite{flensberg} for a nonresonant Josephson current
$J(\varphi)=J_c^{(\lambda)}\sin\varphi$,
$J_c^{(\lambda)}=\exp(-2\lambda^2)J_c$ (where $J_c$ is the critical
current in the absence of electron-vibron interaction, $\lambda=0$).
Later, in Ref.~\onlinecite{zazunovphonon}, it was demonstrated that
in the case $U_C=0$ and $\Delta_0\rightarrow\infty$ an analogous
result,
\begin{eqnarray} \label{Res Jos Curr}
&&J_r(\varphi)=J_r^{(\lambda)}\sin(\varphi/2){\rm
sgn}[\cos(\varphi/2)],
\\
&&J_r^{(\lambda)}=\left(\frac{e\Gamma_0}{\hbar}\right)\exp(-\lambda^2)
,\nonumber
\end{eqnarray}
holds also for the resonant
($\varepsilon_0=0,\Gamma_0=\Gamma_L=\Gamma_R$) current. Notice the
extra factor of 2 in the exponent for the nonresonant critical current.

The renormalization of the bare level width is hard to detect in an
experiment (since one would have to vary the strength of the
electron-vibron interaction). What experimental manifestation of the
Franck-Condon blockade can one then look for? For normal transport
the answer is that with an increase of temperature the lifting of
the blockade is accompanied by a nonmonotonic temperature behavior
of the conductance \cite{krivetemp, krivenature}. An analogous
behavior has been predicted \cite{parafilo} for superconducting
transport, where it is the critical current that reveals an
anomalous $T$-dependence. The characteristic temperature which
determines the peak in a plot of $J_c(T)$ vs. $T$ is  determined by
the polaronic energy shift $E_p\simeq \lambda^2\hbar\omega_0$. For
$T\lesssim E_p$ the critical current  increases with  temperature
(in the regime of temperature-enhanced Josephson coupling) while for
$T\gg E_p$ the current scales as $1/T$ due to a partial cancelation
of Andreev levels contributions. For moderately strong
electron-vibron interactions, $\lambda\gtrsim1$ the crossover from
low-$T$ regime to $1/T$-scaling looks like a ``resonant" enhancement
of the critical current at $T\sim E_p$ (see Fig. 6).

For soft vibrons, $\omega_0\rightarrow0$, the slowly vibrating QD is
always able to change its equilibrium  position
($\langle\hat{x}\rangle=0$) in order to minimize the total energy.
If one neglects electron-electron interactions ($U_C=0$) the total
energy of the weak link can be readily evaluated in the
quasiclassical approximation if the dimensionless operator
$\hat{x}=(b+b^{\dag})/\sqrt{2}$ in Eq.~(\ref{QD HAm}) is replaced by
the classical variable $x_c$. Then the total energy $E_t$ consists
of two terms: (i) the elastic energy, and (ii) the energy of the
filled Andreev level
\begin{equation} \label{Energ}
E_t=\frac{\hbar\omega_0}{2}x^2_c-\sqrt{(\varepsilon_0+\varepsilon_ix_c)^2-\Gamma_0^2\cos^2(\varphi/2)}
\end{equation}
(for simplicity we consider here a symmetric junction). It is easy
to see that for coupling strengths such that
$\varepsilon_i^2\geq\hbar\omega_0\Gamma_0$, a condition which is
always fulfilled in the considered limit $\omega_0\rightarrow0$, the
energy minimum corresponds to a shifted QD position, $x_c\neq0$. In
an effectively asymmetric junction the resonant ($\varepsilon_0=0$)
current is suppressed \cite{parafilo} so that
\begin{equation} \label{Supp curr}
J(\varphi)=J_c\sin\varphi \,,\quad
J_c=\frac{e\omega_0}{2}\left(\frac{\Gamma_0}{\varepsilon_i}\right)^2.
\end{equation}
Unlike the exponential (in the electron-vibron interaction strength)
suppression of the critical current induced by zero-point
fluctuations of the QD coordinate ($\hbar\omega_0\gg\Gamma_0$), soft
vibrons give rise to a power-like (polaronic) suppression of the
form $J\propto \lambda^{-2}$. At finite temperatures thermally
excited vibron polarons (excitations in the state $x_c\neq0$) tend
to shift the QD towards its spatially symmetric position. This means
that $|x_c(T)|<|x_c(0)|$, which implies that at low temperatures the
current grows with an increase of temperature. The crossover from
the regime of a temperature enhanced supercurrent to a standard
$1/T$ scaling of the critical current occurs abruptly at $T\simeq
T_p$ \cite{parafilo}. Therefore, both the Franck-Condon blockade of
the supercurrent and the polaronic effects on the Josephson current
are manifested in an anomalous (nonmonotonic) temperature behavior
of the critical current (see Fig.~6). Estimations show
\cite{parafilo} that for already existing transport experiments on
suspended single wall carbon nanotubes (see e.g. \cite{sapmaz}) the
polaronic temperature is in the range $T_p\sim(1-10)$~K, which makes
the observation of polaronic effects in carbon nanotube-based SNS
junctions a feasible experiment.

\begin{figure}
\vspace{0.cm} \centerline {\includegraphics[width=10cm]{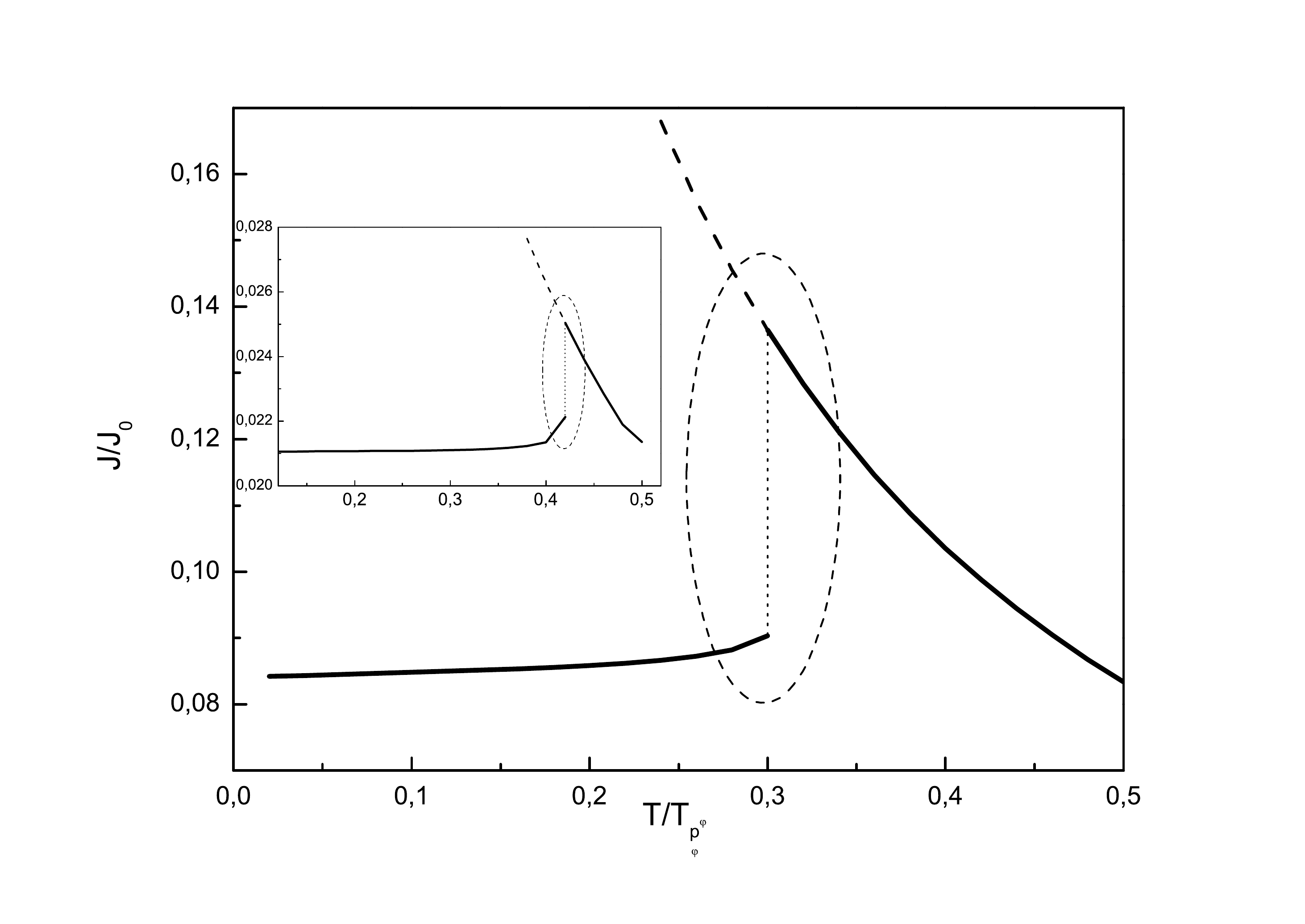}}
\vspace*{-0.5 cm} \caption{Temperature dependence of the Josephson
current for $\varphi=1$, $\hbar\omega_0/\varepsilon_i=0.25$, the
value of $\hbar\omega_0\Gamma_0/\varepsilon_i^2=0.2$ (inset -
$\hbar\omega_0\Gamma_0/\varepsilon_i^2=0.05$). Reprinted with
permission from \cite{parafilo}, A.~V.~Parafilo {\em et al.},
(unpublished). }
\end{figure}

Finally, we discuss  the influence of a charging energy $U_C$ on the
polaronic effects on the Josephson current. To this end we first
note that the electron-vibron interaction renormalizes the
electron-electron correlation energy so that $U_C \rightarrow
U_C^{\rm eff} = U_C-2\lambda^2\hbar\omega_0$ and hence diminishes
the strength of the interaction \cite{flensberg}. It is physically
obvious that as long as $U_C^{\rm eff}\lesssim\Gamma$ the effects of
a finite charging energy are negligible small. When $U^{\rm
eff}_C>\Gamma$ correlations split the energy level $\varepsilon_0$
and for $U^{\rm eff}_C\gg\Gamma$ the conditions for resonant
tunneling of Cooper pairs can not be satisfied. If the
electron-electron correlations are so strong that $U_C^{\rm
eff}\gg\Delta_0$ they additionally suppress the critical current by
a factor $\Gamma/\Delta_0\ll1$. If $E_p\ll U_C^{\rm
eff}\lesssim\Delta_0$ the charging energy does changes the value of
the low-$T$ critical current but it can not influence the predicted
anomalous temperature dependence of the current at $T\simeq E_p$.
What happenes if $U_C^{\rm eff}\sim E_p$ is an open question, which
needs further investigation.

\section{Electro - and Spintro - Mechanics of Magnetic shuttle devices} \label{c}

In this Section we will explore new functionalities that
emerge when nanomechanical devices are partly or completely made of
magnetic materials. The possibility of magnetic ordering
brings new degrees of freedom into play in addition to the
electronic and mechanical ones considered so far, opening up an
exciting perspective towards utilising magneto-electro-mechanical
transduction for a large variety of applications. Device dimensions
in the nanometer range mean that a number of mesoscopic phenomena in
the electronic, magnetic and mechanical subsystems can be used for
quantum coherent manipulations. In comparison with the
electromechanics of the nanodevices considered above the prominent
role of the electronic spin in addition to the electric charge
should be taken into account.

The ability to manipulate and control spins via electrical \cite{9,
10, 11} magnetic \cite{12} and optical \cite{13} means has generated
numerous applications in metrology \cite{14} in recent years. A
promising alternative method for spin manipulation employs
a mechanical resonator
coupled to the magnetic dipole moment of the spin(s), a method which
could enable scalable quantum information architectures \cite{15}
and sensitive nanoscale magnetometry \cite{16, 17, 18}. Magnetic
resonance force microscopy (MRFM) was suggested as a means to
improve spin detection to the level of a single spin and thus enable
three dimensional imaging of macromolecules with atomic resolution.
In this technique a single spin, driven by a resonant microwave
magnetic field interacts with a ferromagnetic particle. If the
ferromagnetic particle is attached to a cantilever tip, the spin
changes the cantilever vibration parameters \cite{21}. The
possibility to detect \cite{21} and monitor the coherent dynamics of
a single spin mechanically \cite{22} has been demonstrated
experimentally. Several theoretical suggestions concerning the
possibility to test single-spin dynamics through an electronic
transport measurement were made recently \cite{24, 25, 26, 27}.
Complementary studies of the mechanics of a resonator coupled to
spin degrees of freedom by detecting the spin dynamics and
relaxation were suggested in \cite{24,25,26,27,28,29,30,31} and
carried out in \cite{32}. Electronic spin-orbit interaction in
suspended nanowires was shown to be an efficient tool for detection
and cooling of bending-mode nanovibrations as well as for
manipulation of spin qubit and mechanical quantum vibrations
\cite{33, 34, 35}.

An obvious modification of the nano-electro-mechanics of magnetic
shuttle devices originates from the spin-splitting of electronic
energy levels, which results in the known phenomenon of
spin-dependent tunneling. Spin-controlled nano-electro-mechanics
which originates from spin-controlled transport of electric charge
in magnetic NEM systems is represented by number of new
magneto-electro-mechanical phenomena.

Qualitatively new opportunities appear when magnetic nanomechanical
devices are used. They have to do with the effect of the
short-ranged magnetic exchange interaction between the spin of
electrons and magnetic parts of the device. In this case the spin of
the electron rather than its electrical charge can be the main
source of the mechanical force acting on movable parts of the
device. This leads to new physics compared with the usual
electromechanics of non-magnetic devices, for which we use the term
spintro-mechanics. In particular it becomes possible for a movable
central island to shuttle magnetization between two magnetic leads
even without any charge transport between the leads. The result of
such a mechanical transportation of magnetization is a magnetic
coupling between nanomagnets with a strength and sign that are
mechanically tunable.

In this Section we will review some early results that involve the
phenomena mentioned above. These only amount to a first step in the
exploration of new opportunities caused by the interrelation between
charge, spin and mechanics on a nanometer length scale.

\subsection{ Spin-controlled shuttling of electric charge} \label{w}

By manipulating the interaction between the spin of electrons and
external magnetic fields and/or the internal interaction in magnetic
materials, spin-controlled nanoelectromechanics may be achieved.

A new functional principle --- spin-dependent shuttling of electrons
--- for low magnetic field sensing purposes was proposed by Gorelik
{\em et al.} in Ref.~\onlinecite{r88}. This principle may lead to a giant
magnetoresistance effect in  external magnetic fields as low as
1-10~Oe in a magnetic shuttle device if magnets with highly
spin-polarized electrons (half metals \cite{r83, r84, r85, r86,
r87}) are used as leads in a magnetic shuttle device. The key idea
is to use the external magnetic field to manipulate the spin of
shuttled electrons rather than the magnetization of the leads. Since
the electron spends a relatively long time on the shuttle, where it
is decoupled from the magnetic environment, even a weak magnetic can
rotate its spin by a significant angle. Such a rotation allows the
spin of an electron that has been loaded onto the shuttle from a
spin-polarized source electrode to be reoriented in order to allow
the electron finally to tunnel from the shuttle to the (differently)
spin-polarized drain lead. In this way the shuttle serves as a very
sensitive ``magnetoresistor" device. The model employed in
Ref.~\onlinecite{r88} assumes that the source and drain are fully polarized
in opposite directions. A mechanically movable quantum dot
(described by a time-dependent displacement $x(t)$), where a single
energy level is available for electrons, performs driven harmonic
oscillations between the leads. The external magnetic field, $H$, is
perpendicular to the orientations of the magnetization in both leads
and to the direction of the mechanical motion.

The spin-dependent part of the Hamiltonian is specified as
\begin{equation} \label{Spin trans ham}
H_{\rm
magn}(t)=J(t)(a^{\dag}_{\uparrow}a_{\uparrow}-a^{\dag}_{\downarrow}a_{\downarrow})
-\frac{g\mu
H}{2}(a^{\dag}_{\uparrow}a_{\downarrow}+a^{\dag}_{\downarrow}a_{\uparrow}),
\end{equation}
where $J(t)=J_R(t)-J_L(t)$, $J_{L(R)}(t)$ are the exchange
interactions between the on-grain electron and the left(right) lead,
$g$ is the gyromagnetic ratio and $\mu$ is the Bohr magneton. The
proper Liouville-von Neumann equation for the density matrix is
analyzed and an average electrical current is calculated for the
case of large bias voltage.

In the limit of weak exchange interaction, $J_{max}\ll \mu H$ one
may neglect the influence of the magnetic leads on the on-dot
electron spin dynamics. The resulting current is
\begin{equation} \label{Mech assi curr}
I=\frac{e \omega_0}{\pi}\frac{\sin^2(\vartheta/2)\tanh(w/4)}{\sin^2
(\vartheta/2)+\tanh^2(w/4)}
\end{equation}
where $w$ is the total tunneling probability during the contact time
$t_0$, while $\vartheta\sim\pi g\mu H/\hbar\omega_0$ is the rotation
angle of the spin during the ``free-motion" time.

The theory \cite{r88} predicts oscillations in the magnetoresistance
of the magnetic shuttle device with a period $\Delta H_p$, which is
determined from the equation $\hbar\omega_0=g \mu (1+w)\Delta H_p$.
The physical meaning of this relation is simple: every time when
$\omega_0/\Omega=n+1/2$ ($\Omega=g \mu H/\hbar$ is the spin
precession frequency in a magnetic field) the shuttled electron is
able to flip fully its spin to remove the ``spin-blockade" of
tunneling between spin polarized leads having their magnetization in
opposite directions. This effect can be used for measuring the
mechanical frequency thus providing dc spectroscopy of
nanomechanical vibrations.

Spin-dependent shuttling of electrons as discussed above is a
property of non-interacting electrons, in the sense that tunneling
of different electrons into (and out of) the dot are independent
events. The Coulomb blockade phenomenon adds a strong correlation of
tunneling events, preventing fluctuations in the occupation of
electronic states on the dot. This effect crucially changes the
physics of spin-dependent tunneling in a magnetic NEM device. One of
the
remarkable consequences is 
the Coulomb promotion of spin-dependent tunneling predicted in
Ref.~\onlinecite{1}. In this work a strong voltage dependence of the
spin-flip relaxation rate on a quantum dot was demonstrated. Such
relaxation, being very sensitive to the occupation of spin-up and
spin-down states on the dot, can be controlled by the Coulomb
blockade phenomenon. It was shown in Ref.~\onlinecite{1} that by lifting
the Coulomb blockade one stimulates occupation of both spin-up and
spin-down states thus suppressing spin-flip relaxation on the dot.
In magnetic devices with highly spin-polarized electrons electronic
spin-flip can be the only mechanism providing charge transport
between oppositely magnetized leads. In this case the onset of
Coulomb blockade, by increasing the spin-flip relaxation rate,
stimulates charge transport through a magnetic SET device (Coulomb
promotion of spin-dependent tunneling). Spin-flip relaxation
qualitatively also modifies the noise characteristics of
spin-dependent single-electron transport. In Refs.~\onlinecite{2, 3} it was
shown that the low-frequency shot noise in such structures diverges
as the spin relaxation rate goes to zero. This effect provides an
efficient tool for spectroscopy of extremely slow spin-flip
relaxation in quantum dots. Mechanical transportation of a
spin-polarized dot in a magnetic shuttle device provides new
opportunities for studying spin-flip relaxation in quantum dots. The
reason can be traced to a spin-blockade of the mechanically aided
shuttle current that occurs in devices with highly polarized and
colinearly magnetized leads. As was shown in Ref.~\onlinecite{4} the
above effect results in giant peaks in the shot-noise spectral
function, wherein the peak heights are only limited by the rates of
electronic spin flips. This enables a nanomechanical spectroscopy of
rare spin-flip events, allowing spin-flip relaxation times as long
as $10~\mu$s to be detected.

The spin-dependence of electronic tunneling in magnetic NEM devices
permits an external magnetic field to be used for manipulating not
only electric transport but also the mechanical performance of the
device. This was demonstrated in Refs.~\onlinecite{s90, 5}. A theory of
the quantum coherent dynamics of mechanical vibrations, electron
charge and spin was formulated and the possibility to trigger a
shuttle instability by a relatively weak magnetic field was
demonstrated. It was shown that the strength of the magnetic field
required to control nanomechanical vibrations decreases with an
increasing tunnel resistance of the device and can be as low as 10
Oe for giga-ohm tunnel structures.

A new type of nanoelectromechanical self excitation caused entirely
by the spin splitting of electronic energy levels in an external
magnetic field was predicted in Ref.~\onlinecite{6} for a suspended
nanowire, where mechanical motion in a magnetic field induces an
electromotive coupling between electronic and vibrational degrees of
freedom. It was shown that a strong correlation between the
occupancy of the spin-split electronic energy levels in the nanowire
and the velocity of flexural nanowire vibrations provides energy
supply from the source of DC current, flowing through the wire, to
the mechanical vibrations thus making possible stable,
self-supporting bending vibrations. Estimations made in
Ref.~\onlinecite{6} show that in a realistic case the vibration amplitude
of a suspended carbon nanotube (CNT) of the order of 10~nm can be
achieved if magnetic field of 10~T is applied.

\subsection{Spintro-mechanics of magnetic shuttle device}

New phenomena, qualitatively different from the electromechanics of
nonmagnetic shuttle systems, may appear in magnetic shuttle devices
in a situation when short-range magnetic exchange forces become
comparable in strength to the long-range electrostatic forces
between the charged elements of the device \cite{6}.  There is
convincing evidence that the exchange field can be several tesla at
a distance of a few nanometers from the surface of a ferromagnet
\cite{radic4, radic5, radic6, radic9}. Because of the exponential
decay of the field this means that the force experienced by a
single-electron spin in the vicinity of magnetic electrodes can be
very large. These spin-dependent exchange forces can lead to various
``spintro-mechanical" phenomena.

Mechanical effects produced by a long-range electrostatic force and
short-ranged exchange forces on a movable quantum dot are
illustrated in Fig.~7.
\begin{figure}
\vspace{0.cm} \centerline {\includegraphics[width=8cm]{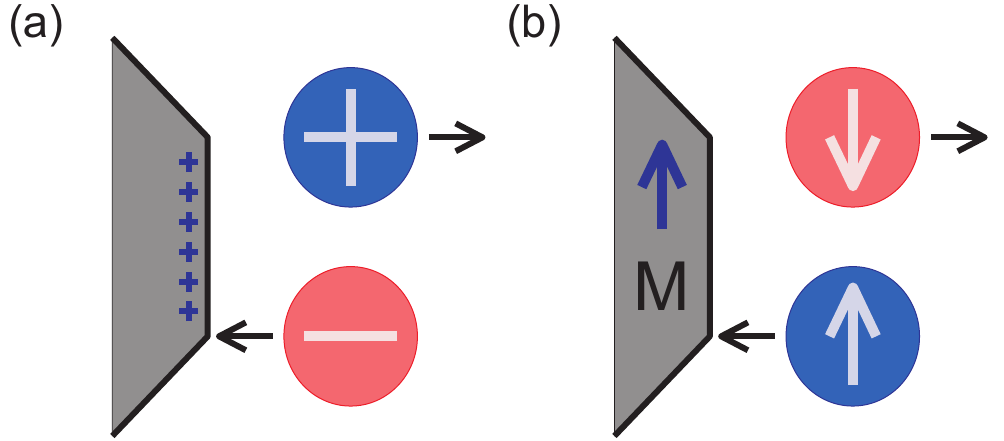}}
\vspace*{-0. cm} \caption{A movable quantum dot in a magnetic
shuttle device can be displaced in response to two types of force:
(a) a long-range electrostatic force causing an electromechanical
response if the dot has a net charge, and (b) a short-rang magnetic
exchange force leading to ``spintromechanical" response if the dot
has a net magnetization (spin). The direction of the force and
displacements depends on the relative signs of the charge and
magnetization, respectively. Reprinted with permission from
\cite{7}, R. I. Shekhter {\em et al.},  {Phys. Rev. B} {\bf 86},
 100404 (2012). $\copyright$ 2012, American Physical Society.}
\end{figure}
The electrostatic force acting on the dot, placed in the vicinity of a
charged electrode (Fig.~7(a)), is determined by the electric charge
accumulated on the dot. In contrast, the exchange force induced by a
neighboring magnet depends on the net spin accumulated on the dot.
While the electrostatic force changes its direction if the electric
charge on the dot changes its sign, the spin-dependent exchange
force is insensitive to the electric charge but it changes direction
if the electronic spin projection changes its sign. A very important
difference between the two forces is that the electrostatic force
changes only as a result of injection of additional electrons into
(out of) the dot while the spintronic force can be changed due to
the electron spin dynamics even for a fixed number of electrons on
the dot (as is the case if the dot and the leads are insulators). In
this case interesting opportunities arise from the possibility of
transducing the dynamical variations of electronic spin (induced,
e.g., by magnetic or microwave fields) to mechanical displacements
in the NEM device. In Ref.~\onlinecite{7} a particular spintromechanical
effect was discussed -- a giant spin-filtering of the electron
current (flowing through the device) induced by the formation of
what we shall call a ``spin-polaronic state".

The Hamiltonian that describes the magnetic nanomechanical SET
device in Ref.~\onlinecite{7} has the standard form (its spin-dependent
part depends now on the mechanical displacement of the dot). Hence
$H=H_{lead}+H_{tunnel}+H_{dot}$, where
$H_{leads}=\sum_{k,\sigma,s}\varepsilon_{ks\sigma}a^{\dag}_{ks\sigma}a_{ks\sigma}$
describes electrons (labeled by wave vector $k$ and spin
$\sigma=\uparrow,\downarrow$) in the two leads ($s=L, R$). Electron
tunneling between the leads and the dot is modeled as
\begin{equation} \label{Spintro Tunn}
H_{tunnel}=\sum_{k,\sigma, s}T_s(x)a^{\dag}_{k
s\sigma}c_{\sigma}+H.c.
\end{equation}
where the matrix elements $T_s(x)=T^{(0)}_s\exp(\mp x/l_t)$ ($l_t$
is the characteristic tunneling length) depend on the dot position
$x$. The Hamiltonian of the movable single-level dot is
\begin{equation} \label{Spintro dot}
H_{dot}=\hbar\omega_0b^{\dag}b+\sum_{\sigma}n_{\sigma}[\varepsilon_0-{\rm sgn}(\sigma)J(x)]+U_Cn_{\uparrow}n_{\downarrow},
\end{equation}
where ${\rm sgn}(\uparrow,\downarrow)=\pm1$, $U_C$ is the Coulomb
energy associated with double occupancy of the dot and the
eigenvalues of the electron number operators $n_{\sigma}$ is $0$ or
$1$. The position dependent magnitude $J(x)$ of the spin dependent
shift of the electronic energy level on the dot is due to the
exchange interaction with the magnetic leads. Here we expand $J(x)$
to linear order in $x$ so that $J(x)=J^{(0)}+j x$ and without loss
of generality assume that $J^{(0)}=0$.

\begin{figure}
\vspace{0.cm} \centerline {\includegraphics[width=6cm]{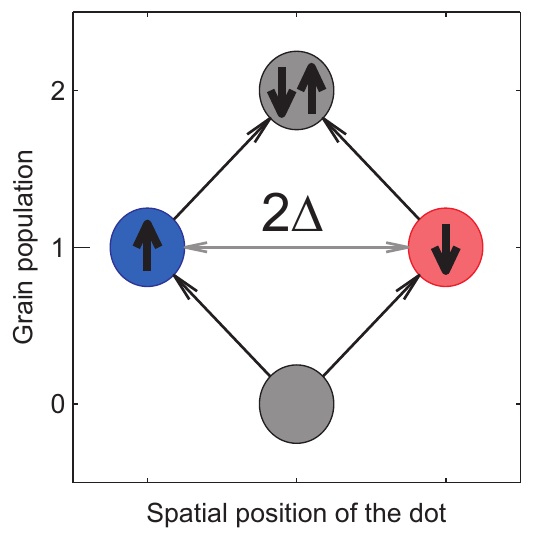}}
\vspace*{-0. cm} \caption{Diagram showing how the equilibrium
position of the movable dot depends on its net charge and spin. The
difference in spatial displacements discriminates transport through
a singly occupied dot with respect to the electron spin. Reprinted
with permission from \cite{7}, R. I. Shekhter {\em et al.}, {Phys.
Rev. B} {\bf 86},
 100404 (2012). $\copyright$ 2012, American Physical Society.}
\end{figure}

The modification of the exchange force, caused by changing the spin
accumulated on the dot, shifts the equilibrium position of the dot
with respect to the magnetic leads of the device. Since the electron
tunneling matrix element is exponentially sensitive to the position
of the dot with respect to the source and drain electrodes one
expects a strong spin-dependent renormalization of the tunneling
probability, which exponentially discriminates between the
contributions to the total electrical current from electrons with
different spins. This spatial separation of dots with opposite spins
is illustrated in Fig.~8. While changing the population of spin-up
and spin-down levels on the dot (by changing e.g. the bias voltage
applied to the device) one shifts the spatial position $x$ of the
dot with respect to the source/drain leads. It is important that the
Coulomb blockade phenomenon prevents simultaneous population of both
spin states.
If the Coulomb blockade
is lifted the two spin states become equally populated with a zero
net spin on the dot, $\textbf{S}=0$. This removes the spin-polaronic
deformation and the dot is situated at the same place as a
non-populated one. In calculations a strong modification of the
vibrational states of the dot, which has to do with a shift of its
equilibrium position, should be taken into account. This results in
a so-called Frank-Condon blockade of electronic tunneling \cite{116,
ratner}. The spintro-mechanical stimulation of a spin-polarized
current and the spin-polaronic Franck-Condon blockade of electronic
tunneling are in competition and their interplay determines a
non-monotonic voltage dependence of the giant spin-filtering effect.

To understand the above effects in more detail consider the
analytical results of Ref.~\onlinecite{7}. A solution of the problem can
be obtained by the standard sequential tunneling approximation and
by solving a Liouville equation for the density matrix for both the
electronic and vibronic subsystems. The spin-up and spin-down
currents can be expressed in terms of transition rates (energy
broadening of the level) and the occupation probabilities for
the dot electronic states.
For simplicity we consider the case of a strongly asymmetric
tunneling device. At low bias voltage and low temperature the
partial spin current is
\begin{equation} \label{Spint curr1}
I_{\sigma}\sim \frac{e
\Gamma_L}{\hbar}\exp\left(\frac{1}{2}\left[\frac{x_0^2}{l_t^2}-\left(\frac{x_0}
{\hbar\omega_0}\right)^2\right]-{\rm sgn}(\sigma)\beta\right),
\end{equation}
where $\beta=x_0^2/\hbar\omega_0l_t$. In the high bias voltage (or
temperature) regime, $max\{eV,T\gg E_p\}$, where the polaronic
blockade is lifted (but double occupancy of the dot is still
prevented by the Coulomb blockade), the current expression takes the
form
\begin{equation} \label{Spintr curr2}
I_{\sigma}\sim \frac{e
\Gamma_L}{\hbar}\exp\left(\left[2n_B+1\right]\frac{x_0^2}{l_t^2}-2\,
{\rm sgn}(\sigma)\beta\right),
\end{equation}
where $n_B$ is Bose-Einstein distribution function. The scale of
the polaronic spin-filtering of the device is determined by the
ratio $\beta$ of the polaronic shift of the equilibrium spatial
position of a spin-polarized dot and the electronic tunneling
length. For typical values of the exchange interaction and
mechanical properties of suspended carbon nanotubes this parameter
is about 1-10. As was shown this is enough for the spin filtering of the
electrical current through the device to be nearly 100 \% efficient.
The temperature and voltage
dependence of the spin-filtering effect is presented in Fig.~9. The
spin filtering effect and the Franck-Condon blockade
\begin{figure}
\vspace{0.cm} \centerline {\includegraphics[width=8cm]{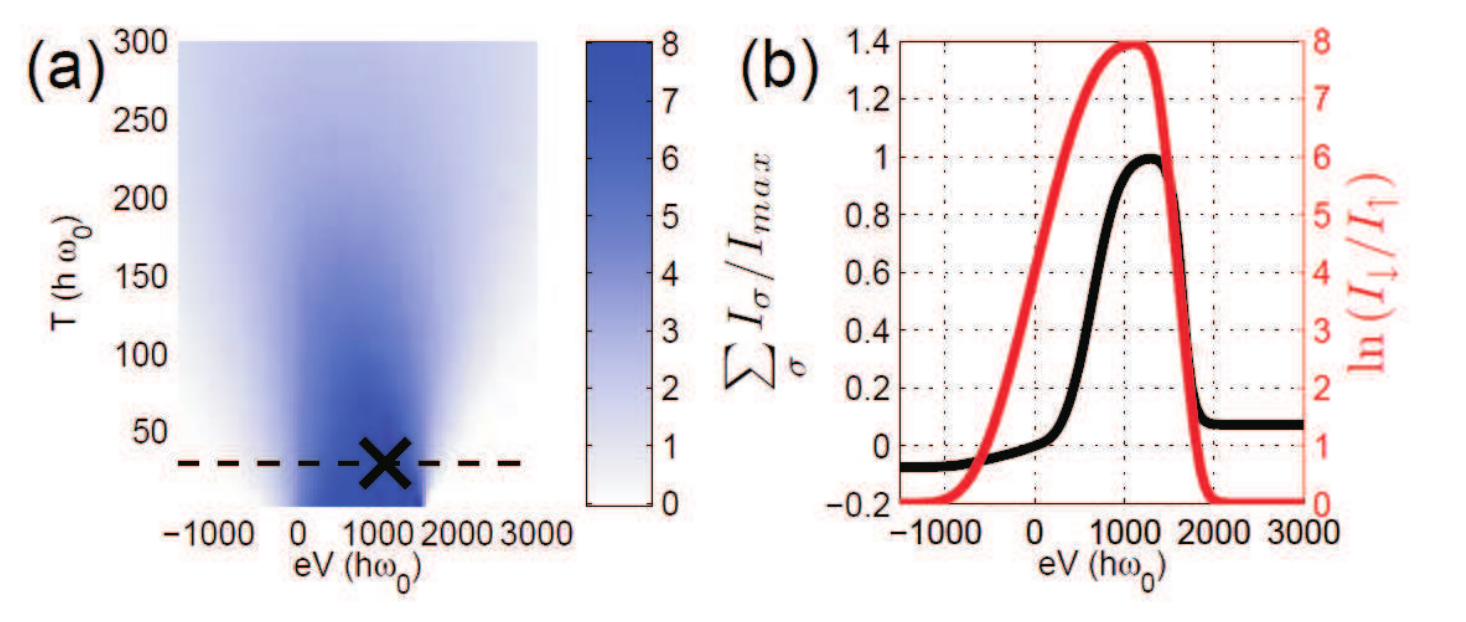}}
\vspace*{-0. cm} \caption{ Spin polarization of the current through
the model NEM-SET device under discussion. Reprinted with permission
from \cite{7}, R. I. Shekhter {\em et al.}, {Phys. Rev. B} {\bf 86},
 100404 (2012). $\copyright$ 2012, American Physical Society.}
\end{figure}
both occur at low voltages and temperatures (on the scale of the
polaronic energy; see Fig.~9 (a)).  An increase of the voltage
applied to the device lifts the Franck-Condon blockade, which
results in an exponential increase of both the current and the
spin-filtering efficiency of the device. This increase is blocked
abruptly at voltages for which the Coulomb blockade is lifted. At
this point a double occupation of the dot results in spin
cancellation and removal of the spin-polaronic segregation. This
leads to an exponential drop of both the total current and the spin
polarization of the tunnel current (Fig.~9~(b)). As one can see in
Fig.~9 prominent spin filtering can be achieved for realistic device
parameters. The temperature of operation of the spin-filtering
device is restricted from above by the Coulomb blockade energy. One
may, however, consider using functionalized nanotubes
\cite{pulkin20} or graphene ribbons \cite{pulkin21} with one or more
nanometer-sized metal or semiconductor nanocrystal attached. This
may provide a Coulomb blockade energy up to a few hundred kelvin,
making spin filtering a high temperature effect \cite{7}.

\subsection{Mechanically assisted magnetic coupling between
nanomagnets}

The mechanical force caused by the exchange interaction represents
only one effect of the coupling of magnetic and mechanical degrees of
freedom in magnetic nanoelectromechanical device. A complementary
effect is the of mechanical transportation of magnetization, which
we are going to discuss in this subsection.

\begin{figure}
\vspace{-0.cm} \includegraphics[width=7cm]{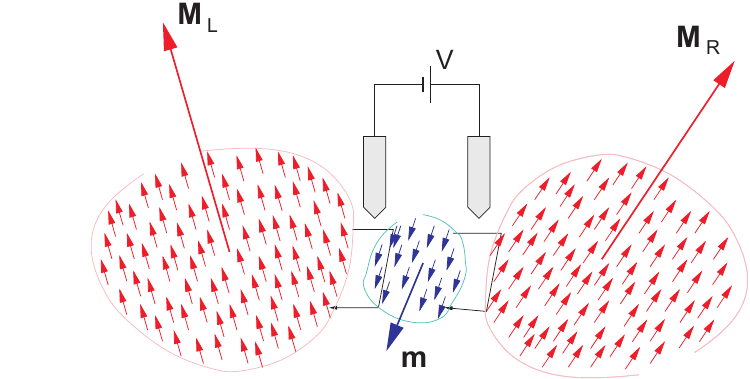}
\vspace*{-0.cm} \caption{Single-domain magnetic grains with magnetic
moments $\textbf{M}_L$ and $\textbf{M}_R$ are coupled via a magnetic
cluster with magnetic moment $\textbf{m}$, the latter being
separated from the grains by insulating layers. The gate electrodes
induce an ac electric field, concentrated in the insulating regions.
This field, by controlling the heights of the tunnel barriers,
affects the exchange magnetic coupling between different components
of the system. Reprinted with permission from \cite{8}, L. Y.
Gorelik et al., {Phys. Rev. Lett.} {\bf 91},
 088301 (2003). $\copyright$ 2003, American Physical Society.}
\end{figure}

In the magnetic shuttle device presented in Fig.~10, a ferromagnetic
dot with total magnetic moment $\textbf{m}$ is able to move between
two magnetic leads, which have total magnetization
$\textbf{M}_{L,R}$. Such a device was suggested in Ref.~\onlinecite{8} in
order to consider the magnetic coupling between the leads (which in
their turn can be small magnets or nanomagnets) produced by a
ferromagnetic shuttle. It is worth to point out that the phenomenon
we are going to discuss here has nothing to do with transferring
electric charge in the device and it is valid also for a device made
of nonconducting material. The main effect, which will be in the focus
of our attention, is the exchange interaction between the
ferromagnetic shuttle (dot) and the magnetic leads. This interaction
decays exponentially when the dot moves away from a lead and hence
it is only important when the dot is close to one of the leads.
During the periodic back-and-forth
motion of the dot this happens during short time intervals near the
turning points of the mechanical motion. An exchange interaction
between the magnetizations of the dot and a lead results in a
rotation of these two magnetization vectors in such a way that the
vector sum is conserved. This is why the result of this rotation can
be viewed as a transfer of some magnetization $\Delta\textbf{m}$
from one ferromagnet to the other. As a result the magnetization of
the dot experiences some rotation around a certain axis. The total
angle $\phi$ of the rotation accumulated during the time when the
dot is magnetically coupled to the lead is an essential parameter
which depends on the mechanical and magnetic characteristics of the
device. The continuation of the mechanical motion breaks the
magnetic coupling of the dot with the first lead but later, as the
dot
approaches
the other magnetic lead an exchange coupling is established with
this second lead with the result that
magnetization which is ``loaded" on the dot from the first lead is
"transferred" to the this second lead.
This is how the transfer of magnetization from one magnetic lead to
another is induced mechanically. The transfer creates an
effective coupling between the magnetizations of the two leads. Such
a non-equilibrium coupling can be efficiently tuned by controlling
the mechanics of the shuttle device. It is particularly interesting
that the sign of the resulting magnetic interaction is determined by
the sign of $\cos (\phi/2)$. Therefore, the mechanically mediated
magnetic interaction can be changed from ferromagnetic to
anti-ferromagnetic by changing the amplitude and the frequency of
mechanical vibrations \cite{8}.

\subsection{Resonance spin-scattering effects. Spin shuttle as a ``mobile quantum impurity".}

The Kondo effect in electron tunneling results from the spin
exchange between electrons in the leads and the island (quantum dot)
that couples the leads and manifests itself as a sharp zero bias
anomaly in the low-temperature tunneling conductance. Many-particle
interactions and the tunneling renormalize the electron spectrum
enabling Kondo resonances  both for odd \cite{gogo} and even
\cite{Pust00,Kikoin01} electron occupations. In the latter case the
Kondo resonance is caused by the singlet-triplet crossover in the
ground state (see \cite{glasko} for a review). In the simplest case
of odd occupancy a cartoon of a quantum well and a schematic Density
of States (DoS) is shown in Fig.~11. For simplicity we consider a
case when the dot is occupied by one electron (as in a SET
transistor). The dot level is not in resonance with the Fermi level
of the leads ($\epsilon_F$), but located at an energy $-E_d$, below
it. The dot is in the Coulomb blockade regime and the corresponding
charging energy is denoted as $E_C$. The resonance spin scattering
results in the formation of a narrow peak in the DoS known as the
Abrikosov-Suhl resonance \cite{Abrikosov, Suhl, Hewson} (see
Fig.~11, right panel). The width of this resonance defines a unique
energy scale, the Kondo temperature
 $T_K$, which determines all thermodynamic and transport properties of the
SET device through a one-parametric scaling \cite{Hewson}.
The width $\Gamma$ of the dot level,  associated with the tunneling
of dot electrons to the continuum of levels in the leads, is assumed
to be smaller
than the charging energy $E_C$, providing a condition for an integer
valency regime. When the shuttle moves between source (S) and drain
(D) (see the lower panel of Fig. 11), both the energy $E_d$ and the
width $\Gamma$ acquire a time dependence. This time dependence
results in a coupling between mechanical and electronic degrees of
freedom. If a source-drain voltage $V_{sd}$ is small enough
($eV_{sd}\ll T_K$) the charge degree of freedom of the shuttle is
frozen out while spin plays a very important role in co-tunneling
processes. Namely, the dot electron's spin can be flipped while the
electron tunnels from the left to the right lead. Thus, the initial
and final states of the quantum impurity can have different spins.
This process is accompanied by simultaneous creation of spin
excitations in the Fermi sea. The many-electron scattering processes
then lead to the formation of an Abrikosov-Suhl resonance. This
resonance can be viewed as a Kondo cloud built up from both
conduction electrons in the leads and a localized electron in the
dot. Since all electrons in the cloud contain information about the
same impurity, they are mutually correlated. Thus, NEM  providing a
coupling between mechanical and electronic degrees of freedom
introduces a powerful tool for manipulation and control of the Kondo
cloud and gives a very promising and efficient mechanism
\color{black} for electromechanical transduction on the nanometer
length scale.

\begin{figure}[t]
\vspace*{0mm}
\includegraphics[angle=0,width=\columnwidth]{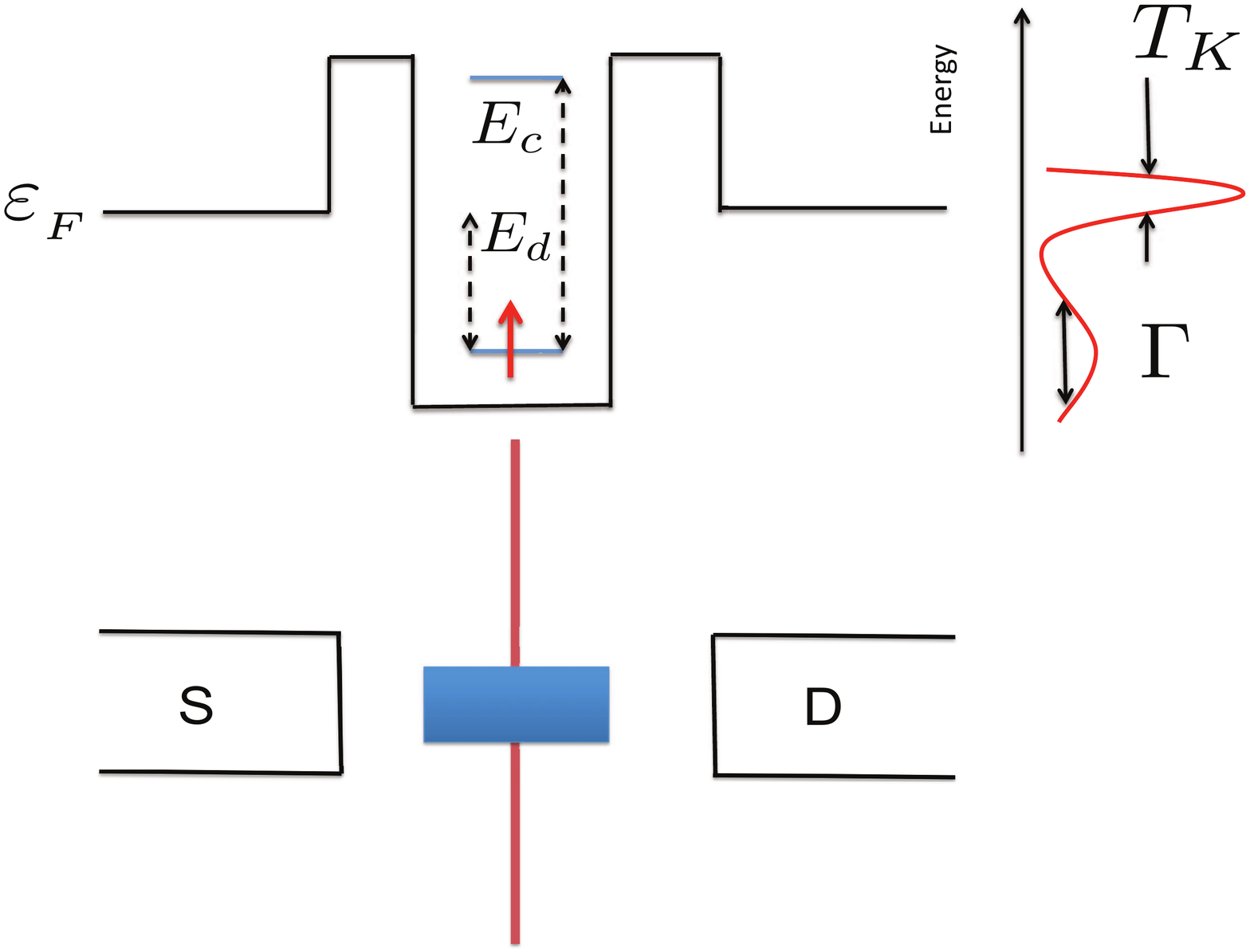}
\vspace*{-2mm} \caption{  Nanomechanical resonator with spin as a
``mobile quantum impurity".} \label{f.1}
\end{figure}

Building on an analogy with the shuttling experiments of Refs. \onlinecite{Kot2004}
 and \onlinecite{erbe}, let us consider a device where an isolated
nanomachined island oscillates between two electrodes (Fig.11, lower
panel). The applied voltage is assumed low enough so that the field
emission of many electrons, which was the main mechanism of
tunneling in those experiments, can be neglected. We emphasize that
the characteristic de Broglie wave length associated with the dot
should be much shorter than typical displacements allowing thus for
a classical treatment of the mechanical motion of the nano-particle.
The condition $\hbar\omega_0\ll T_K$,  necessary to eliminate
decoherence effects, requires for e.g. planar quantum dots with the
Kondo temperature $T_K\gtrsim 100$~mK, the condition
$\omega_0\lesssim 1$~GHz for oscillation frequencies to hold; this
frequency range is experimentally feasible \cite{erbe,Kot2004}. The
shuttling island is then to be considered as a ``mobile quantum
impurity", and transport experiments will detect the influence of
mechanical motion on the differential conductance. If the dot is
small enough, then the Coulomb blockade guarantees the single
electron tunneling or cotunneling regime, which is necessary for the
realization of the Kondo effect \cite{GR,glasko}.  Cotunneling
is accompanied by a change of spin projection in the process of
charging/discharging of the shuttle and therefore is closely related
to the spin/charge pumping problem \cite{brouw98}.

A generic Hamiltonian for describing the resonance spin-scattering
effects is given by the Anderson model,
\begin{eqnarray}
&&H_0 =\sum_{k,\alpha}
\varepsilon_{k\sigma,\alpha}c^\dagger_{k\sigma,\alpha}c_{k\sigma,\alpha}
+\sum_{i\sigma}[E_{d}-e Ex] d^{\dagger}_{i\sigma} d_{i\sigma} + E_c
n^2 \nonumber
\\
&& H_{tun}=
\sum_{ik\sigma,\alpha}T^{(i)}_{\alpha}(x)[c^\dagger_{k\sigma,\alpha}d_{i\sigma}
+ H.c],\label{ham1}
\end{eqnarray}
where $c^\dagger_{k\sigma}$, $d^\dag_{i\sigma}$ create an electron
in the lead $\alpha$$=$$L$$,$$R$, or the dot level
$\varepsilon_{i=1,2}$, respectively,
 $n=\sum_{i\sigma}d^\dagger_{i\sigma}d_{i\sigma}$, ${ E}$ is the electric field between the leads. The tunnelling matrix element
$T^{(i)}_{L,R}(x)=T^{(i,0)}_{L,R}\exp[\mp x(t)/l_t]$, depends
exponentially on the ratio of the time-dependent displacement $x(t)$
and the electronic tunnelling length $l_t$.  The time-dependent
Kondo Hamiltonian can be obtained from it by applying a
time-dependant Schrieffer-Wolff transformation \cite{SW}:
\begin{eqnarray}
H_K=\sum_{k\alpha\sigma, k'\alpha'\sigma'}{\cal
J}_{\alpha\alpha'}(t)[\vec{\sigma}_{\sigma\sigma'}\vec{S}+
\frac{1}{4}\delta_{\sigma\sigma'}]c^\dagger_{k\sigma,\alpha}c_{k'\sigma',\alpha'}
\label{ham2}
\end{eqnarray}
where ${\cal
J}_{\alpha,\alpha'}(t)=\sqrt{\Gamma_{\alpha}(t)\Gamma_{\alpha'}(t)/(\pi\rho_0
E_d(t))}$ and $\vec
S=\frac{1}{2}d_\sigma^\dagger\vec\sigma_{\sigma\sigma'} d_\sigma'$,
$\Gamma_\alpha(t)=2\pi \rho_0 |T_\alpha(x(t))|^2$ are level widths
due to tunneling to the left and right leads.

As long as the nano-particle is not subject to an external
time-dependent electric field, the Kondo temperature is given by
$T_K^0=D_0\exp\left[-(\pi E_C)/(8\Gamma_0)\right]$ (for simplicity
we assumed that $\Gamma_L(0)=\Gamma_R(0)=\Gamma_0$; $D_0$ plays the
role of effective bandwidth). As the nano-particle moves
adiabatically, $\hbar \omega_0\ll \Gamma_0$, the decoherence effects
are small provided $\hbar\omega_0\ll T_K^0$.
\begin{figure}[h]
  \includegraphics[width=60mm,angle=0]{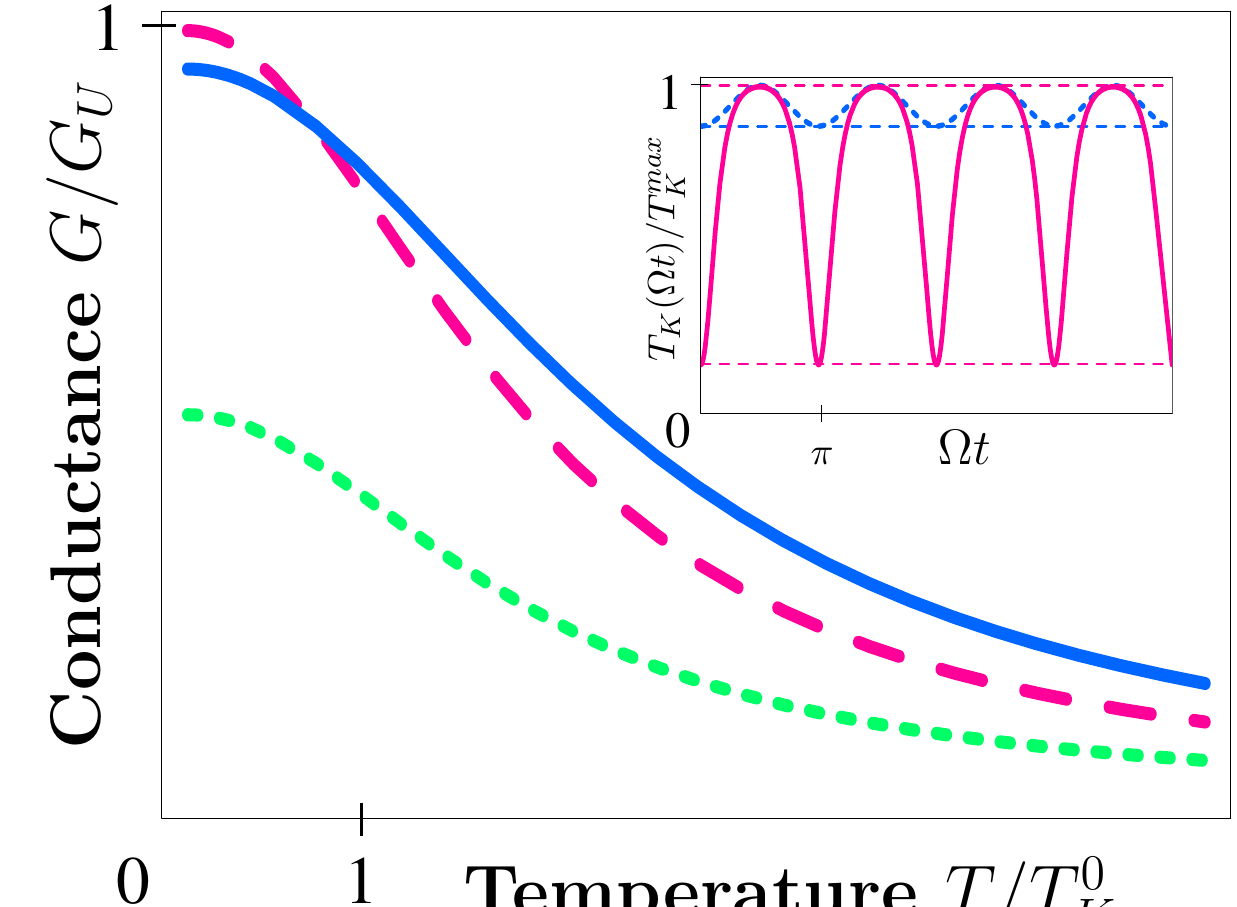}
  \caption{ Differential conductance $G$ of a Kondo shuttle for which $\Gamma_0$$/$$U$$=$$0.4$.
  The solid line denotes $G$ for a shuttle with $\Gamma_L$$=$$\Gamma_R$, $A$$=$$l_t$,
 the dashed line shows $G$ for a static nano-island with $\Gamma_L=\Gamma_R$, $A$$=$$0$,
 the dotted line gives $G$ for $\Gamma_L$$/$$\Gamma_R$$=$$0.5$, $A$$=$$0$.
 The inset shows the  temporal oscillations (here $\Omega\equiv\omega_0$) of $T_K$ for small
 $A$$=$$0.05$$\,l_t$ (dotted line) and large $A$$=$$2.5$$\,l_t$ (solid line) shuttling amplitudes.
 Reprinted with permission from \cite{kis06},
M. N. Kiselev {\em et al.}, {Phys. Rev. B} {\bf 74},
 233403 (2006). $\copyright$ 2006, American Physical Society.
  }\label{fig:sh2}
\end{figure}

Let us first assume a temperature regime $T\gg T_K$ (weak coupling).
In this case we can build a perturbation theory controlled by the
small parameter $\rho_0 {\cal J}(t)\ln[D_0/T]<1$ assuming time as an
external parameter. The series of perturbation theory can be summed
up by means of a renormalization group procedure \cite{Hewson}. As a
result, the Kondo temperature becomes oscillating in time:
\begin{eqnarray}
T_K(t)=D(t) \exp\left[-\frac{\pi
E_C}{8\Gamma_0\cosh(2x(t)/l_t)}\right].\label{ftk}
\end{eqnarray}
Neglecting the weak time-dependence of the effective bandwidth
$D(t)\approx D_0$, we arrive at the following expression for the
time-averaged Kondo temperature:
\begin{eqnarray}
\langle T_K\rangle=T_K^0 \bigg\langle \exp\left[\frac{\pi
E_C}{4\Gamma_0}\frac{\sinh^2 (x(t)/l_t)}{1+2\sinh^2
(x(t)/l_t)}\right]\bigg\rangle. \label{tkon1}
\end{eqnarray}
Here $\langle$$...$$\rangle$ denotes averaging over the period of
the mechanical oscillation. The expression (\ref{tkon1}) acquires an
especially transparent form when the amplitude of the mechanical
vibrations $A$ is small: $A\lesssim l_t$. In this case the Kondo
temperature can be written as $\langle T_K \rangle = T_K^0
\exp(-2W)$, with the Debye-Waller-like exponent $W =-\pi E_C \langle
x^2(t)\rangle)/(8\Gamma_0l_t^2)$, giving rise to the enhancement of
the static Kondo temperature.

The zero bias anomaly (ZBA) in the tunneling conductance is given by
\begin{eqnarray}
G(T) =\frac{3\pi^2}{8}G_0\Bigg\langle\frac{4\Gamma_L(t)\Gamma_R(t)}
{(\Gamma_L(t)+\Gamma_R(t))^2}\frac{1}{[\ln(T/T_K(t))]^2}\Bigg\rangle,
\end{eqnarray}
where $G_0=e^2/h$ is a unitary conductance. Although the central
position of the island is most favorable for the Breit-Wigner (BW)
resonance ($\Gamma_L =\Gamma_R$), it corresponds to the minimal
width of the Abrikosov-Suhl resonance. The turning points correspond
to the maximum of the Kondo temperature given by the equation
(\ref{ftk}) while the system is away from the BW resonance. These
two competing effects lead to the effective enhancement of $G$ at
high temperatures (see Fig. 12).
\begin{figure}[t]
\vspace*{-0mm}
\includegraphics[angle=0,width=7cm]{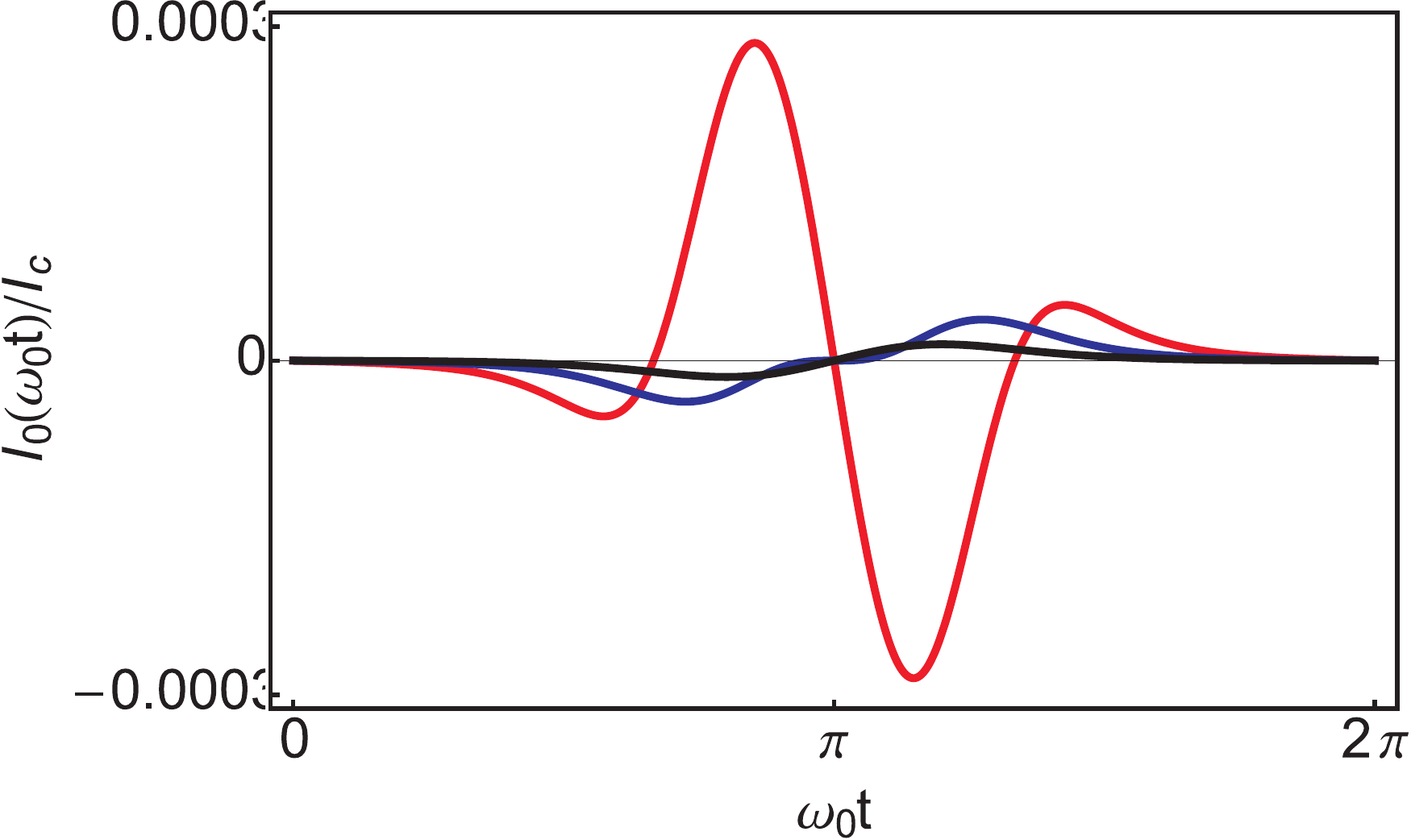}
\vspace*{-0.mm} \caption{Time dependence of the current $I_0$ for
different values of asymmetry parameter $u=x_0/l_t$. Here red, blue
and black curves correspond to $u=0.5; 1.0; 1.5;$. For all three
curves shuttle oscillates with amplitude $x_{max}=l_t$,
$\hbar\omega_0/(T_K)^{\color{black}min\color{black}}=10^{-3}$,
$|eV_{\rm bias}|/T_K^{\color{black}min\color{black}}=g\mu_B
B/T_K^{\color{black}min\color{black}}=0.1$ with $T_K^{(0)}=2 K$,
$l_t/L=10^{-4}$. Reprinted with permission from \cite{kis12}, M. N.
Kiselev {\em et al.} Phys. Rev. Lett. {\bf 110}, 066804 (2013).
$\copyright$ 2013, American Physical Society.} \label{ff.1}
\end{figure}

Summarizing, it was shown in \cite{kis06} that Kondo shuttling in a
NEM-SET device increases the Kondo temperature due to the asymmetry
of coupling at the turning points compared to at the central
position of the island. As a result, the enhancement of the
differential conductance in the weak coupling regime can be
interpreted as a pre-cursor of strong electron-electron correlations
appearing due to formation of the Kondo cloud.

Let us consider now the strong coupling regime, $T\ll T_K$. The
current through the system subject to a constant source-drain bias
$V_{sd}$ can be separated in two parts: a dc current associated with
a time-dependent dc conductance and an ac current related to the
periodic motion of the shuttle. For both currents Kondo physics
plays an important role. While the dc current is mostly responsible
for the frequency shift, the ac current gives an access to the
dynamics of the Kondo cloud and provides information about the
kinetics of its formation. In order to evaluate both contributions
to the total current we rotate the electronic states in the leads in
such a way that only one combination of the wave functions is
coupled to the quantum impurity. The cotunneling Hamiltonian may be
rationalized by means of the Glazman-Raikh rotation, parametrized by
the angle $\vartheta_t$ defined by the relation $\tan \vartheta_t =
\sqrt{|\Gamma_R(t)/\Gamma_L(t)|}$.

Both the ac and dc contributions to the current can be calculated by
using Nozi\`ere's Fermi-liquid theory (see \cite{Nozieres74} for
details). The ac contribution, associated with the time dependence
of the Friedel phase $\delta_{\sigma}$ \cite{kis12}, is given by
\begin{equation}\label{3.15}
\bar I_{ac}(t) = \frac{\dot {x}(t)}{l_t} \frac{e
E_C}{8\Gamma_0}\cdot\frac{e V_{\rm sd}}{ T_K(t)}\cdot \frac{
\tanh\left(\frac{2[x(t)-x_0]}{l_t}\right)}{
\cosh^{2}\left(\frac{2[x(t)-x_0]}{l_t}\right)}\\
\end{equation}
($\exp(4 x_0/l_t)=\Gamma_R(0)/\Gamma_L(0)$) and the  ``ohmic" dc
contribution is fully defined by the adiabatic time-dependence of
the Glazman-Raikh angle
\begin{equation}\label{3.16}
\bar I_{DC}(t) = G_0 V_{sd} \sin^2
2\vartheta_t\sum_\sigma\sin^2\delta_\sigma
\end{equation}
As a result, the ac contribution to the total current can be
considered as a first non-adiabatic correction:
\begin{equation}\label{4.5}
I_{tot} = I_{ad}(x(t))-\dot x \frac{dI_{ad}}{dx}\frac{\hbar \pi
E_C}{16 \Gamma_0  T^{(0)}_K}
\end{equation}
where  $I_{ad}= 2\cdot  G_0\cdot V_{sd}\cosh^{-2}(2[x(t)-x_0]/l_t)$
and $T_K^{(0)}$ is the Kondo temperature at the equilibrium
position. The small correction to the adiabatic current  in
(\ref{4.5}) may be considered as a first term in the  expansion over
the small  non adiabatic parameter \color{black} $\omega_0\tau\ll
1$\color{black}, where $\tau$ is the retardation time associated
with the inertia of the Kondo cloud. Using such an interpretation
one gets $\tau= \hbar \pi E_C/(16 \Gamma_0  T^{(0)}_K)$.

Equation (\ref{4.5}) allows one  to obtain information about the
dynamics of the Kondo clouds from an analysis of an experimental
investigation of the  mechanical vibrations. \color{black} The
retardation time associated with the dynamics of the Kondo cloud is
parametrically large compared with the time of formation of the
Kondo cloud $\tau_K=\hbar/T_K$ and can be measured owing to a small
deviation from adiabaticity. \color{black} Also we would like to
emphasize a supersensitivity of the quality factor to a change of
the equilibrium position of the shuttle characterized by the
parameter $u$ (see Fig.~13). The influence of strong coupling
between mechanical and electronic degrees of freedom on the
mechanical quality factor has been considered in \cite{kis12}. It
has been shown that both suppression $Q>Q_0$ and enhancement $Q<Q_0$
of the dissipation of nanomechanical vibrations (depending on
external parameters and the equilibrium position of the shuttle) can
be stimulated by  Kondo tunneling.  The latter case demonstrates the
potential for a Kondo induced electromechanical instability.

Summarizing,  we emphasize that the Kondo phenomenon in single
electron tunneling \color{black} gives a very promising and
efficient mechanism \color{black} for electromechanical transduction
on a nanometer length scale. Measuring the nanomechanical response
on Kondo-transport in a nanomechanical single-electron device
enables one to study the kinetics of the formation of
Kondo-screening and offers a new approach for studying
nonequilibrium Kondo phenomena.  The Kondo effect provides a
possibility for super high tunability of the mechanical dissipation
as well as super sensitive detection of mechanical displacement.

\section{Experimental observation of electron shuttling}

Electron shuttling was theoretically predicted to occur in
mechanically soft mesoscopic systems about 15 year ago
\cite{shuttleprl}. Since then there has been a steadily increasing
interest in studying this nonequilibrium electromechanical
phenomenon from both theoretical and experimental points of view.
Each year the technical capability to fabricate shuttle-like devices
improves. On the experimental side there are two main directions in
the study of mechanically mediated electron transport: (i) electron
shuttling in NEMS with \textbf{intrinsic} electromechanical
coupling, and (ii) electron transfer caused by an \textbf{external}
excitation of mechanical motion. Here we briefly review several
recent publications, which have claimed to observe electron
shuttling. We start with the nanoelectromechanical devices based on
vibrating cantilevers.

Externally driven nanomechanical shuttles have been designed in
Refs.~\onlinecite{erbe1, erbe}. In these experiments a
nanomechanical pendulum was fabricated on a Si-on-insulator
substrate using electron and optical lithography. A metal island was
placed on a clapper, which could vibrate between source and drain
electrodes (see Fig.~14). The pendulum was excited by applying an ac
voltage between two gates on the left- and right-hand sides of the
clapper. The observed tunneling source-drain current was strongly
dependent on the frequency of the exciting signal having pronounced
maxima at the eigenfrequencies of the mechanical modes. This fact
signalizes a shuttling mechanism of electron transfer at typical
shuttle frequencies of about 100~MHz. The measured average dc
current at 4.2~K corresponds to $0.11\pm0.001$ electrons per cycle
of mechanical motion. Both a theoretical analysis and numerical
simulations showed that a large portion of the voltage also acts on
the island.

\begin{figure}
\vspace{0.cm} \centerline {\includegraphics[width=7cm]{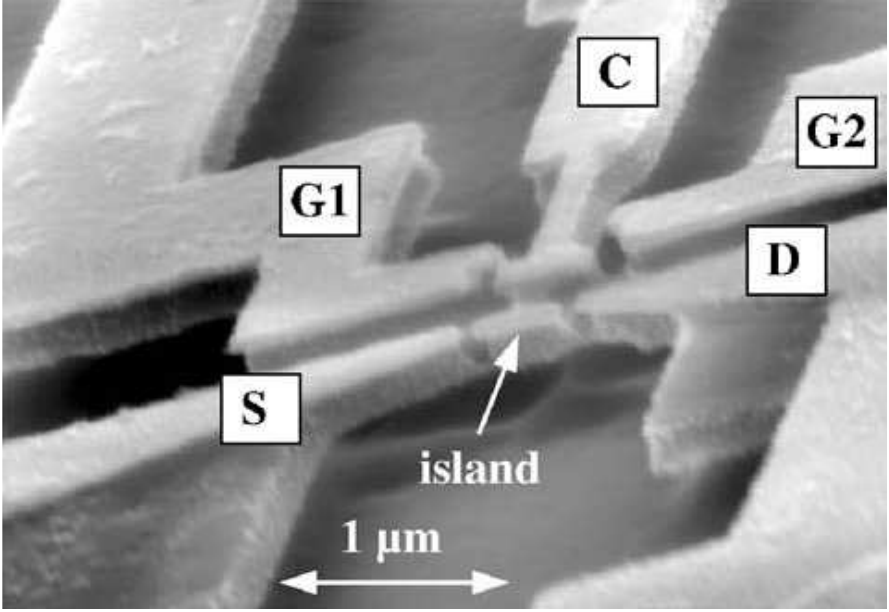}}
\vspace*{-0. cm} \caption{Electron micrograph of a ``quantum bell":
The pendulum is clamped on the upper side of the structure. It can
be set into motion by an ac-power, which is applied to the gates on
the left- and right-hand side (G1 and G2) of the clapper (C).
Electron transport is then observed from source (S) to drain (D)
through the island on top of the clapper. The island is electrically
isolated from the rest of the clapper, which is grounded. Reprinted
with permission from \cite{erbe1}, A.~Erbe {\em et al.}, Appl. Phys.
Lett. {\bf 73}, 3751 (1998). $\copyright$ 1998, American Institute
of Physics.}
\end{figure}

A very important modification of the setup in Fig.~14 was presented
in Ref.~\onlinecite{sheibe}. There a silicon cantilever is part of a
mechanical system of coupled resonators, which is a construction that makes
it possible to drive the shuttle mechanically with a minimal
destructive influence from the actuation dynamics on the shuttle
itself. This is achieved by a clever design that minimizes the
electrical coupling between the driving part of the device (either a
magnetomotively driven, doubly clamped beam resonator, or a
capacitively coupled remote cantilever) and the driven part (the
cantilever that carries the shuttle on its tip). In principle, systems of this
type can be used for studies of shuttle transport
through superconducting and magnetic systems.

In Ref.~\onlinecite{beam} the role of nanocantilever was played by a
semiconductor nanowire. In was shown that under certain conditions
the constant electron beam produced by a scanning electron
microscope can excite self-sustained mechanical oscillations of
semiconducting SiC nanowire. The nanowire plays the role of
mechanical resonator and may be represented by an $RC$ circuit
element (where $R$ is the nanowire resistance and $C$ is the
capacitance between the nanowire end and its environment or some
electrode placed near the resonator). The periodic electrostatic
force, which depends on the charge of the wire, acts on the wire due
to variations in capacitance. The charge on the nanowire is also a
time-dependent function. Discharging occurs due to $RC$ relaxation
accompanied by the drift of electrons to the tungsten tip, on which
the wire is attached. The charging is provided by the electron beam
(for details see \cite{beam}). The semiconductor nanowire starts to
oscillate and goes to the stationary cycle. Thus in \cite{beam} a
new type of electromechanical coupling was studied.


\begin{figure}
\vspace{0.cm} \centerline {\includegraphics[width=7cm]{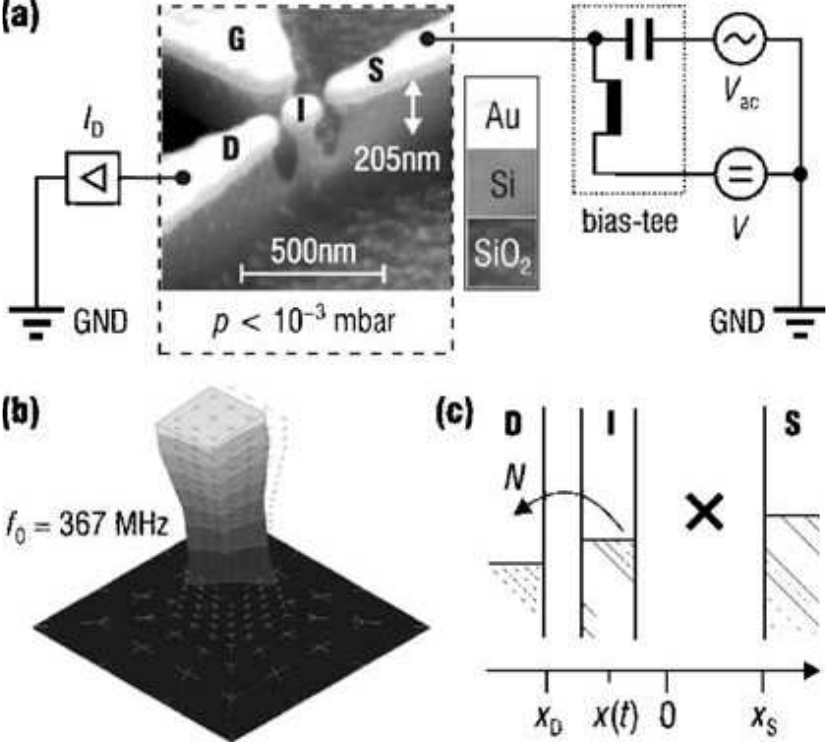}}
\vspace*{-0. cm} \caption{(a) SEM micrograph and experimental
circuitry of the silicon nanopillar studied in Ref.~\onlinecite{nm}:
At the source (S), an ac signal, $V_{ac}$, is applied with a
superimposed dc bias $V$. The net current $I_D$ is detected at the
drain (D) with a current amplifier. The third electrode (G) is
floating. (b) Finite-element simulation of the base oscillation mode
that compiles for the nanopillar to $f_0=5367$~MHz. (c) When the
island is deflected toward one electrode, the instantaneous voltage
bias determines the preferred tunneling direction. Co-tunneling is
absent in this case due to an increased distance to the opposite
electrode. Reprinted with permission from \cite{nm}, D.~V.~Scheibe
and R.~H.~Blick, Appl. Phys. Lett {\bf 84}, 4632 (2004).
$\copyright$ 2004, American Institute of Physics. }
\end{figure}

Interesting results on mechanically assisted charge transfer were
obtained in Ref.~\onlinecite{nm} for a device fabricated as a silicon
nanopillar located between source and drain contacts (see Fig.~15).
The device was manufactured in a two step process: nanoscale
lithography using a scanning electron microscope (SEM) and, second,
dry etching in a fluorine reactive ion etcher (RIE). The
lithographically defined gold structure acts both as electrical
current leads and etch mask for the RIE. A simple geometry defined
by SEM consequently results in the freestanding isolating nanopillar
of intrinsic silicon with a conducting metal (Au) island at its top
(see Fig.~15). This island serves as the charge shuttle. The metal
island and the nanopillar are placed in the center of two facing
electrodes. The system is biased by an ac voltage at source, rather
than a sole dc bias, to avoid the dc-self excitation. Application of
an ac-signal excites one of the nanopillar eigenmodes resonantly.
The device was operated at room temperature and the capacitance was
not sufficiently small to realize the Coulomb blockade regime. The
dependencies of the current on bias frequency, as well as on an
additional dc bias, allowing to tune resonances, were measured.

\begin{figure}
\vspace{0.cm} \centerline {\includegraphics[width=5cm]{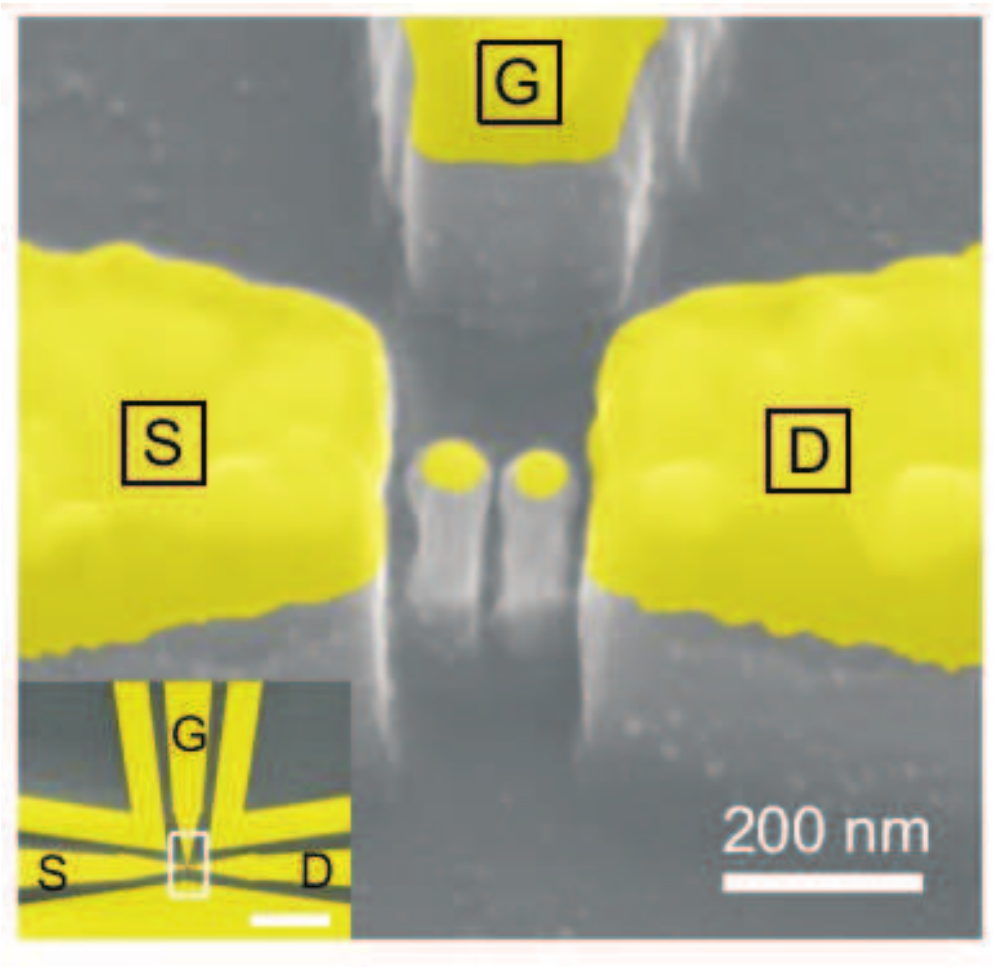}}
\vspace*{-0. cm} \caption{Two coupled electron shuttles realized as
nanopillars. The metallic top layer allows electron exchange with
the source (S) and drain (D) contacts. The scale bar corresponds to
a length of 200~nm. The inset shows a broader view of the
coplanar-waveguide into which the nanopillars are embedded. The
scale bar in the inset is 10~$\mu$m. Reprinted with permission from
\cite{kim}, C. Kim {\em et al.},  {ACS Nano} {\bf 6}, 651 (2012).
$\copyright$ 2012, American Chemical Society.}
\end{figure}

In Refs.~\onlinecite{kim2} and \onlinecite{kim}  electron transport at room temperature
through two nanopillars was considered. The nanopillars act as
shuttles placed in series between source and drain electrodes under
ac/dc excitation (see Fig.~16). The linear size of the island on top of
each pillar was 65~nm with pillar heights of 250~nm and an
inter-pillar distance of 17~nm. At first the I-V
characteristics at room temperature were measured \cite{kim} without
any ac signal. The bias voltage dependence of the current was shown to
be almost linear (except for a small deviation $\propto V^2$ above 1~V).
At low voltages ($<100$~mV) Coulomb blockade
features (a CB staircase) are apparent. From experimental data one finds the
charging energy of the two coupled nanopillars to be $E_C=41$~meV, which is
larger than the (room) temperature equivalent of the experimental setup.
\begin{figure}
\vspace{0.cm} {\includegraphics[width=6cm]{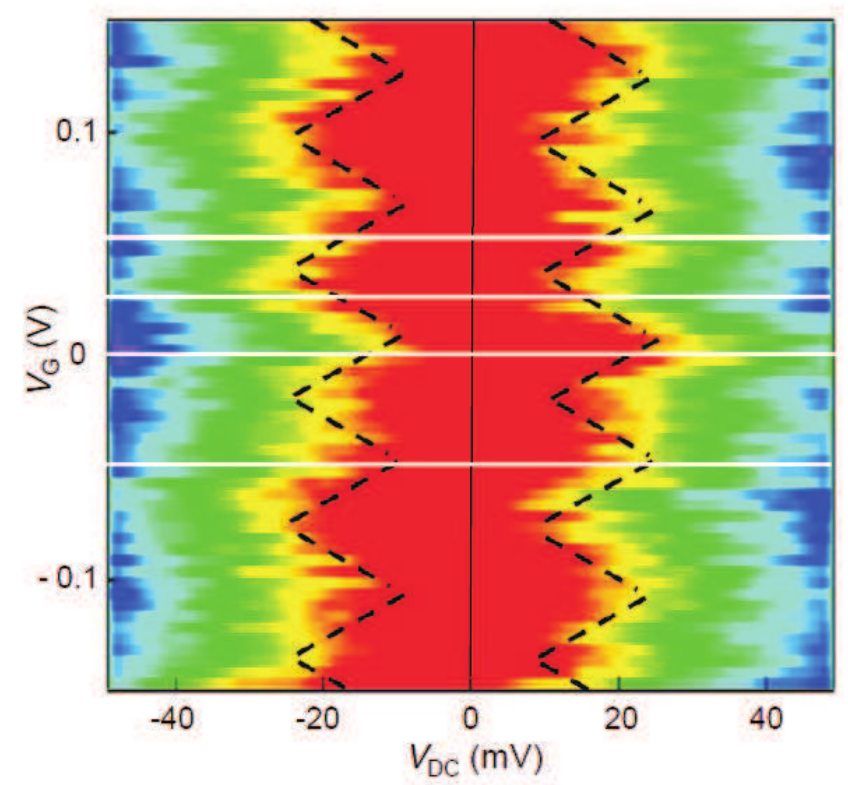}} \vspace*{-0.
cm} \caption{Experimental Coulomb blockade (CB) diamonds traced in
the normalized current in color representation. The lower borders of
the CB regions are represented in red with the Coulomb plateaus
depicted in green. The borders of the CB determined from the theory
plots are marked by dashed black lines as a guide to the eye.
Reprinted with permission from \cite{kim}, C. Kim {\em et al.}, {ACS
Nano} {\bf 6}, 651 (2012). $\copyright$ 2012, American Chemical
Society.}
\end{figure}

The
 dependence of current on gate and bias voltages is manifested in the
Coulomb diamonds (see Fig. 17) in agreement with theoretical
calculations. Here the typical Coulomb blockade staircase
(superimposed on an ohmic response) is smeared due to thermal
broadening and shuttling effects.
\begin{figure}
\vspace{0.cm} \centerline {\includegraphics[width=8cm]{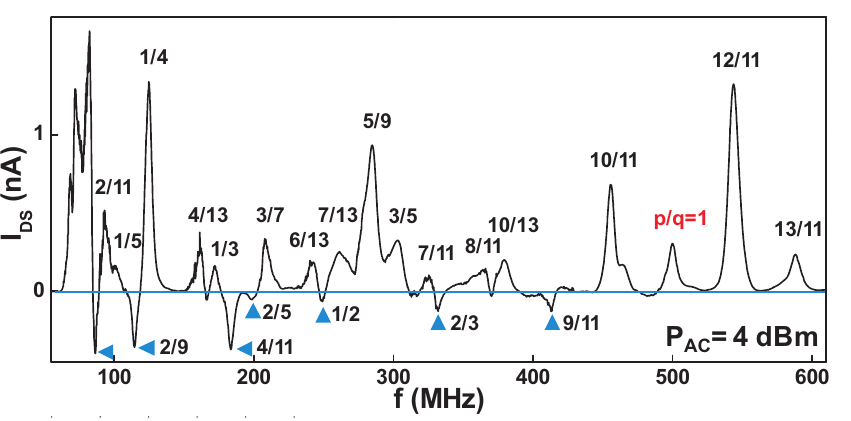}}
\vspace*{-0. cm} \caption{Full frequency sweep of the direct current
through the coupled shuttle revealing the mechanical mode structure
for $V_{DC}=0$. Reprinted with permission from \cite{kim2}, C. Kim
{\em et al.},  {Phys. Rev. Lett.} {\bf 105}, 067204 (2010).
$\copyright$ 2010, American Physical Society.}
\end{figure}

Then, a radio frequency signal (1~MHz - 1~GHz) was fed without any
dc bias added. Due to the alternating-voltage induced ac current one
would expect a zero-average (over an ac period) dc current without
the mechanical subsystem  However, in the considered experiment a
nonzero net current was observed. The authors explained this fact by
the excitation of a mechanical motion of the nanopillars. The effect
of current rectification indicates a dynamical violation of
P-symmetry (symmetry with respect to coordinates reflection
$\textbf{r}\rightarrow -\textbf{r}$) in the system. The direction
and the amplitude of the dc current depends on the ac frequency (see
Fig.~18). The presence of a broad set of resonances indicates the
existence of different mechanical modes of the coupled nanopillars
(Fig.~18).
\begin{figure}
\vspace{0.cm} {\includegraphics[width=6cm]{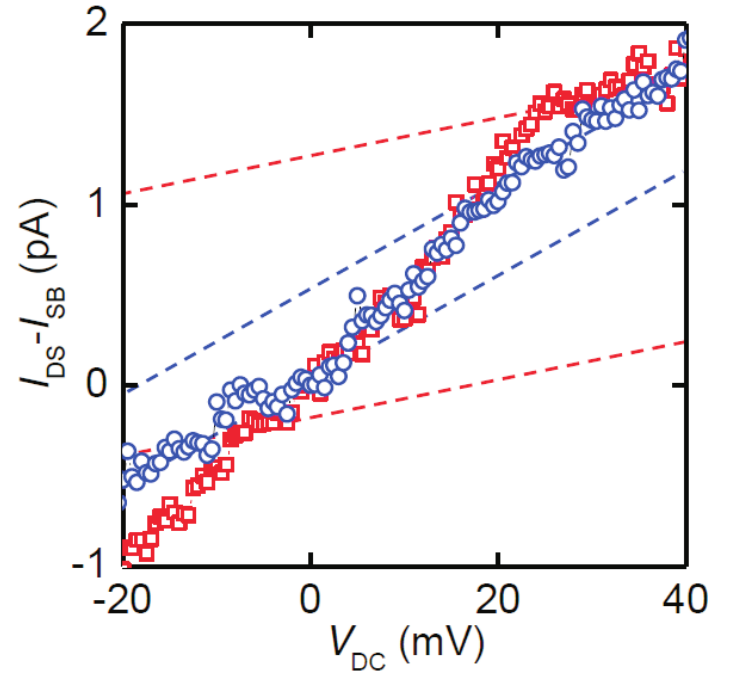}} \vspace*{-0.
cm} \caption{I-V traces for two mechanical modes of the shuttle
device shown in Fig.~16. As can be seen the slopes at the plateaus
(indicated by dashed lines) increase from a shuttling frequency of
285~MHz (red squares) to 500~MHz (blue circles). Reprinted with
permission from \cite{kim}, C. Kim {\em et al.},  {ACS Nano} {\bf
6}, 651 (2012). $\copyright$ 2012, American Chemical Society.}
\end{figure}
Different mechanical modes were investigated by applying a low dc bias voltage.
The observed I-V characteristics (with the zero-bias current
subtracted) is plotted in Fig.~20. The
step-like features in the current-voltage dependence can be
interpreted \cite{kim} as a signature of electron shuttling in the
Coulomb blockade regime.

\begin{figure}
\vspace{0.cm} \centerline {\includegraphics[width=8cm]{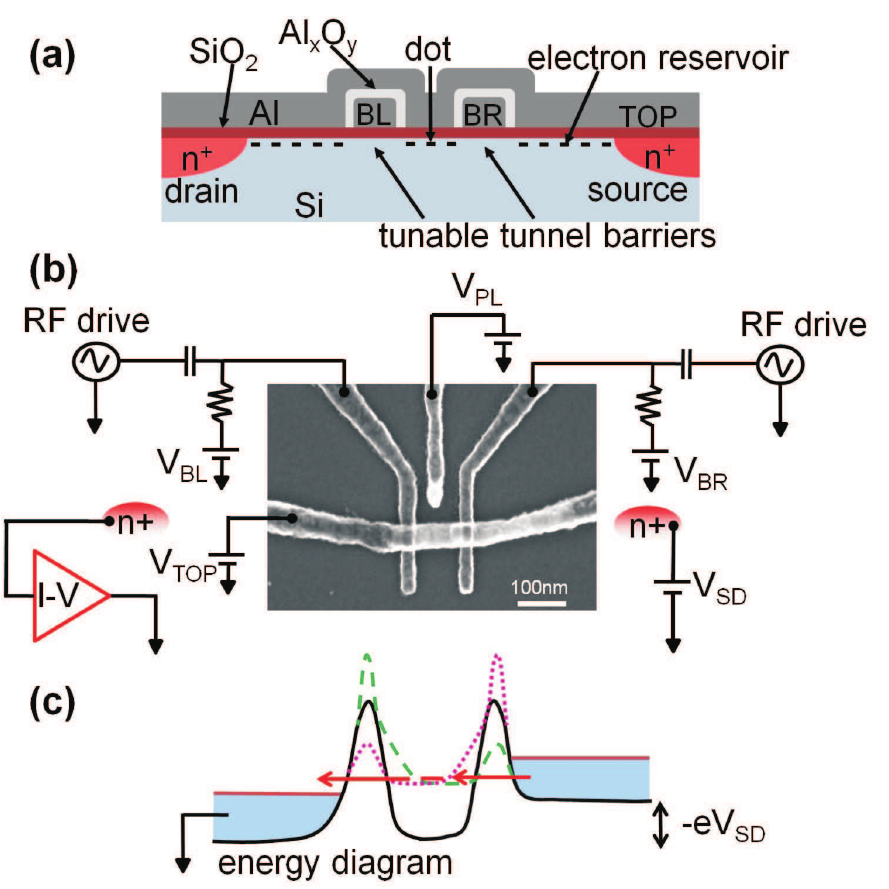}}
\vspace*{-0. cm} \caption{(a) Schematic cross section of a
fabricated silicon metal-oxide-semiconductor (MOS) quantum dot
\cite{Chan}. Two aluminum barrier gates (BL and BR) are below a top
gate (TOP) isolated with Al$_x$O$_y$. The source and drain are
thermally diffused with phosphorus. (b) Scanning electron microscope
image of the device with a simplified measurement setup. (c) Energy
landscape through the dot and lead reservoirs with an illustration
of the electron shuttling. When the sinusoidal ac voltage on BR,
$V_{BR}$ achieves its maximum (dashed green line), an electron
tunnels into the dot from the right electron reservoir. After one
half of an operation period $V_{BL}$ is at its maximum value (dotted
purple line) and the electron tunnels away. Blue regions denotes the
states in the leads occupied by electrons. Reprinted with permission
from \cite{Chan}, K.~W.~Chan {\em et al.}, {Appl. Phys. Lett.} {\bf
98}, 212103 (2011). $\copyright$ 2009, American Institute of
Physics.}
\end{figure}

Another trend in the study of electron shuttles is to mimic shuttling
effects by time dependent tunneling barriers. The authors of
Ref.~\onlinecite{Chan} investigated single-electron ``shuttling" with a silicon
metal-oxide-semiconductor quantum dot at low temperature (300~mK).
The analyzed system represents an electron layer at a
Si/SiO$_2$ interface below an aluminium top gate. This layer forms a
conduction channel between source and drain electrodes (see Fig.~19).
The quantum dot is formed by two additional gates,
which produce tunnel barriers. The effects of mechanical degrees of
freedom of the quantum dots are mimicked here by an ac voltage of frequency
$f_p$ applied to both gates. Periodic variations of the barrier
heights induce periodic effective ``displacements" of the quantum dot
on the nanometer scale. Fig.~21 demonstrates the dependence of the
current on the bias voltage for different frequencies, $I_{dc}=\pm n
e f_p$. The observed results can be explained by a sequential
tunneling model with the electron temperature is a fitting parameter.
Electrons are ``hotter" than the environment due to ac voltage heating.

\begin{figure}
\vspace{0.cm} \centerline {\includegraphics[width=8cm]{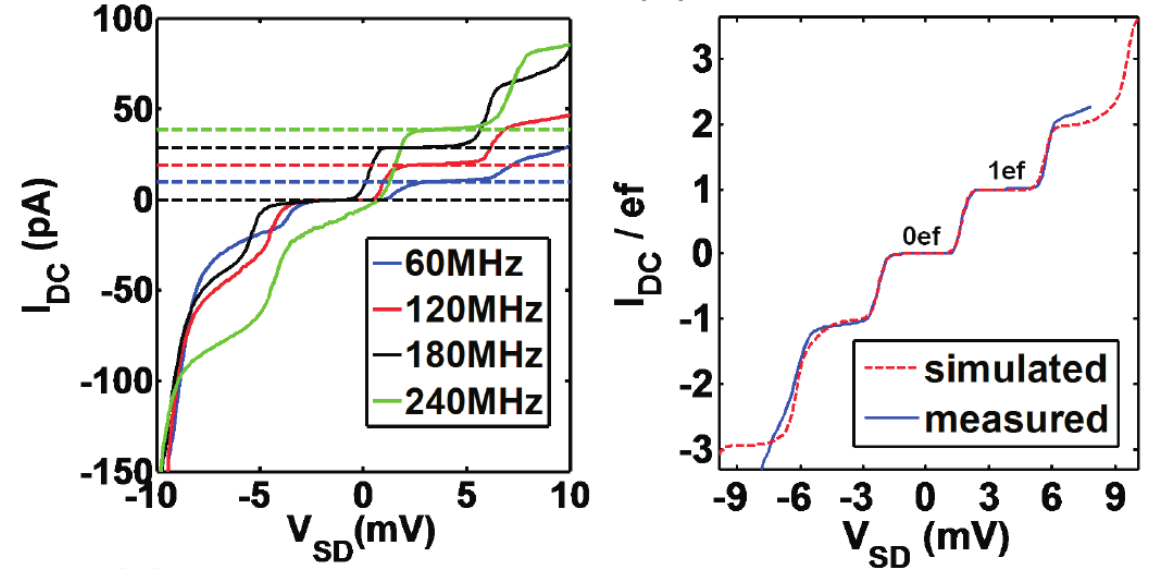}}
\vspace*{-0. cm} \caption{(a) Charge shuttling current measured for
different frequencies. (b) Measured charge shuttling current (solid
blue line) for $f_p=60$~MHz. The measured current is compared with a
simulation (dotted red line) based on a sequential tunneling model
with variable tunneling resistances in the barriers. Reprinted with
permission from \cite{Chan}, K.~W.~Chan {\em et al.}, { Appl. Phys.
Lett.} {\bf 98}, 212103 (2011). $\copyright$ 2009, American
Institute of Physics.}
\end{figure}

An elegant experiment was performed by Park {\em et al.}
\cite{park}, who studied electron transport through a fullerene
molecule placed in the gap between two gold electrodes. The
experimental results (a stair-case-like dependence of current on
bias voltage) are well explained by the process of vibron-assisted
tunneling \cite{braigflens}. The shuttle model was also used for an
explanation of the step-like features on I-V characteristics (see
the discussion in reviews \cite{shuttlerev1, shuttlerev2}).

\begin{figure}
\vspace{0.cm} \centerline {\includegraphics[width=8cm]{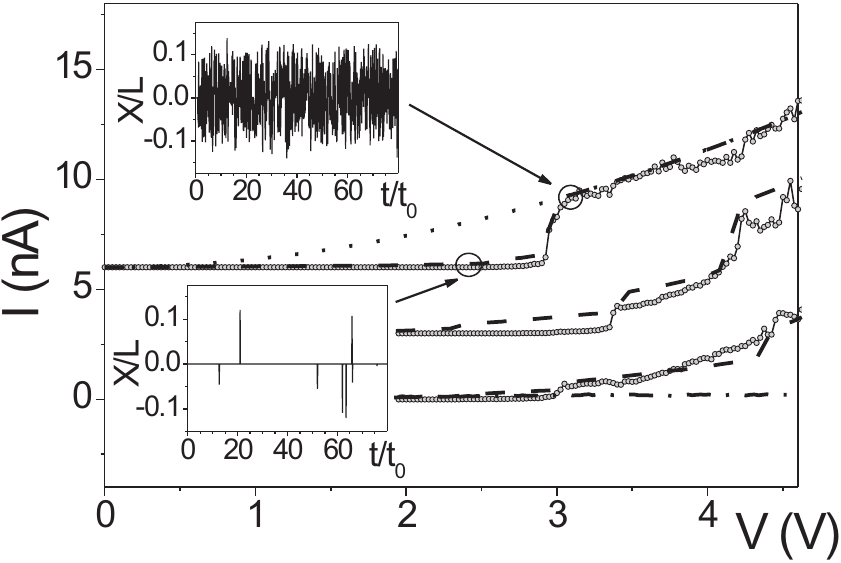}}
\vspace*{-0. cm} \caption{Experimental (symbols) and simulated
(dotted and dashed lines) current-voltage characteristics for
shuttle junctions fabricated A.~V.~Moskalenko {\em et al.}
\cite{Moskalenko}. The dotted line corresponds to oscillations in
the case of zero pinning and the dashed lines to the case of finite
pinning in the system. Insets show the shuttle displacement as a
function of time for two points, one of which is below and the other
is above the transition into the shuttling regime. The leakage
current through a monolayer of octanedithiol molecules is shown by
the dashed-dotted line. Reprinted with permission from
\cite{Moskalenko}, A.~V.~Moskalenko {\em et al.}, {Phys. Rev. B}
{\bf 79}, 241403 (2009). $\copyright$ 2009, American Physical
Society.}
\end{figure}

\begin{figure}
\vspace{0.cm} \centerline {\includegraphics[width=8cm]{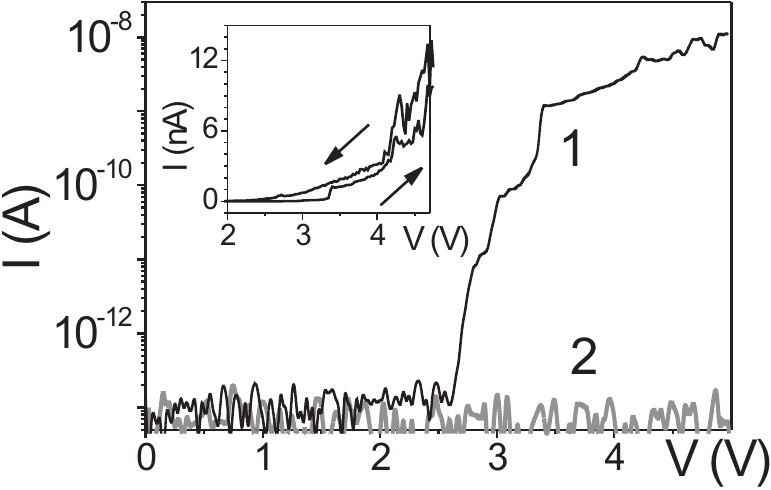}}
\vspace*{-0. cm} \caption{Confirmation that current flows through
the nanoparticle in the shuttling regime (curve 1). If the
nanoparticle is removed from the gap using an AFM tip the result is
a drop in the current through the device of several orders of
magnitude (curve 2). The inset shows hysteretic behavior of I-V
curves obtained for a working shuttle junction in regimes of
increasing and decreasing applied voltage. Reprinted with permission
from \cite{Moskalenko}, A.~V.~Moskalenko {\em et al.}, {Phys. Rev.
B} {\bf 79}, 241403 (2009). $\copyright$ 2009, American Physical
Society.}
\end{figure}
 In Ref.~\onlinecite{Moskalenko} the authors reported shuttling by a
 20~nm gold particle pasted into the gap (10-20~nm) between two
electrodes and attached to them through a monolayer of organic
molecules (1.8 octanedithiol). Experimental I-V characteristics are
presented in Fig.~22. One can see that the theoretical fit using
 shuttle model is in a good agreement with experiment.
At low voltages the current through the device is absent (the nanoparticle
is in a locked state). At high voltages the nanoparticle begins to
vibrate (the quantum dot escapes from the locked state), because more
electrons are transferred to the granule and the electrostatic force
acting on it becomes strong enough to de-pin the shuttle. The onset
of shuttle instability occurs at higher bias voltages than predicted
by a theory assuming frictionless shuttle motion. This discrepancy can
be explained by the binding of the particle to the electrodes
(respectively the value of the threshold voltage is increased).
Experimental data in favor of this assumption is the observation of an
I-V hysteresis loop (see Fig.~23). Without a gold particle the current
through the device is determined by the sequential tunneling of
electrons. Comparison of the experimental data and theoretical
calculations suggests that the average number of electrons that are
involved in the shuttling at voltages of order 3~V is about 20. This
experiment can be interpreted as electron shuttling but the device
is operated in a regime very far from single electron tunneling
(Coulomb blockade regime). Is it possible to fabricate a shuttle-like
device operated in the Coulomb blockade regime by dc bias voltage?
Although this shuttle was not observed yet, the ``road" to its
discovery now is sufficiently clear.

\begin{figure}
\vspace{0.cm} \centerline {\includegraphics[width=8cm]{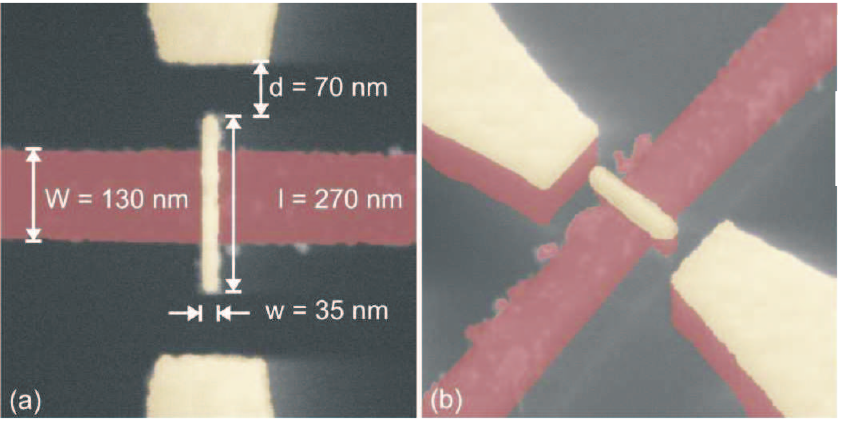}}
\vspace*{-0. cm} \caption{A nanomechanical electron shuttle. (a) SEM
image of a shuttle indicating the dimensions of the gold island
(yellow) and the high stress silicon nitride resonator (red)
suspended above the silicon substrate (grey). Tilted view of gold
island between source and drain. Reprinted with permission from
\cite{self},D.~R.~Koenig and E.~M.~Weig, Appl. Phys. Lett {\bf 101},
213111 (2012).$\copyright$ 2009, American Institute of Physics.}
\end{figure}

In Ref.~\onlinecite{self} the authors claimed to have found a charge shuttle (by the
observation of sustained self-oscillation) operated solely by an
applied dc voltage without external actuation. The island of the
shuttle device in Ref.~\onlinecite{self} is a gold particle with typical
dimensions 35~nm$\times$270~nm$\times$40~nm, placed at the center of
a suspended silicon nitride string (see Fig.~24). Experiments were
performed in helium exchange gas with a pressure of 0.5~mbar in
a helium dewar at $T=4.2$~K. Forty-four samples were studied for
statistical processing. To observe voltage-sustained
self-oscillations the resonant acoustic drive is turned off when the
island is charged. After that the stable charge transport is
observed for almost 2000~s (about $10^{10}$ shuttle cycles). The
reasons for the collapse of self-sustained oscillations
\begin{figure}
\vspace{0.cm} \centerline {\includegraphics[width=7cm]{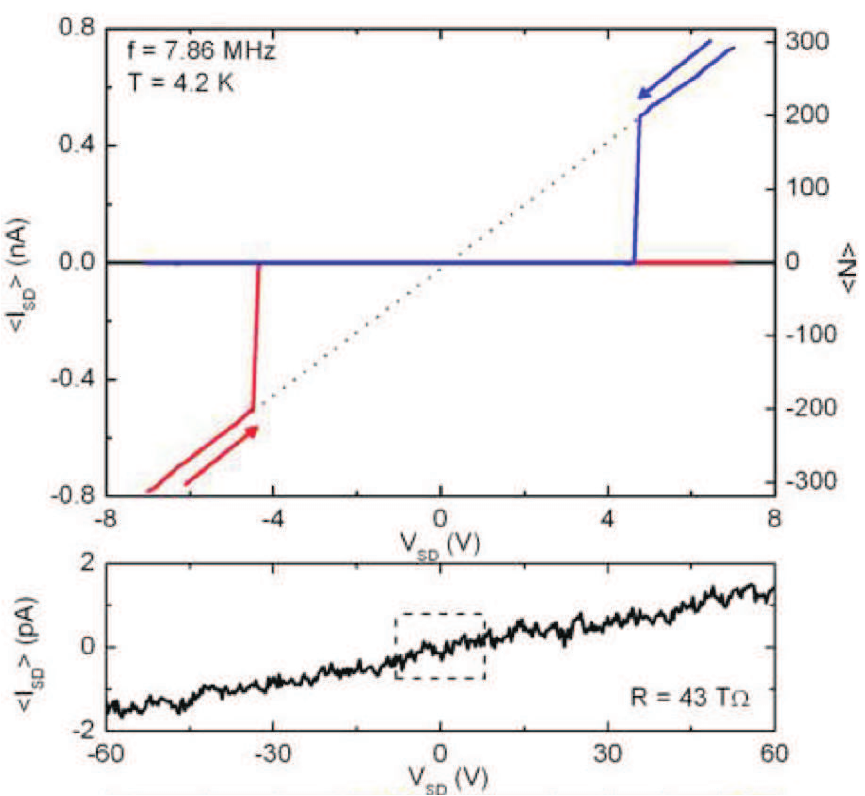}}
\vspace*{-0 cm} \caption{DC voltage-sustained electron shuttling and
background current. (a) Current-voltage curves of voltage-sustained
self-sustained oscillation. Both the blue and the red trace,
corresponding to downward and upward voltage sweeps, respectively,
feature a sharp dissipation threshold. (b) Background current
determined by measuring $I_{SD}$ in the absence of mechanical
shuttling as a function of bias voltage. The dashed box indicates
the voltage range depicted in (a), where the background current is
also shown as a black line. Reprinted with permission from
\cite{self}, D.~R.~Koenig and E.~M.~Weig, Appl. Phys. Lett {\bf
101}, 213111 (2012).$\copyright$ 2009, American Institute of
Physics.}
\end{figure}
may be different: due to impact-induced coupling  to
out-of-plane or torsional motion or wear-induced alternation of
island and electrodes, etc. In the absence of mechanical shuttling
the current is a linear function of bias voltage (see Fig.~25). The
presence of electron shuttling is indicated by a step-like current
dependence. The average number of transferred electrons ($N\simeq200$)
is easily found from the I-V characteristics (see Fig.~25).  The
advantage of the dc-biased self-sustained shuttle current is the
significant drop of external heat load on the system. The described
experiment opens the pathway to observe a Coulomb blockade shuttle
operated solely by a dc bias.

\begin{figure}
\vspace{0.cm} {\includegraphics[width=7cm]{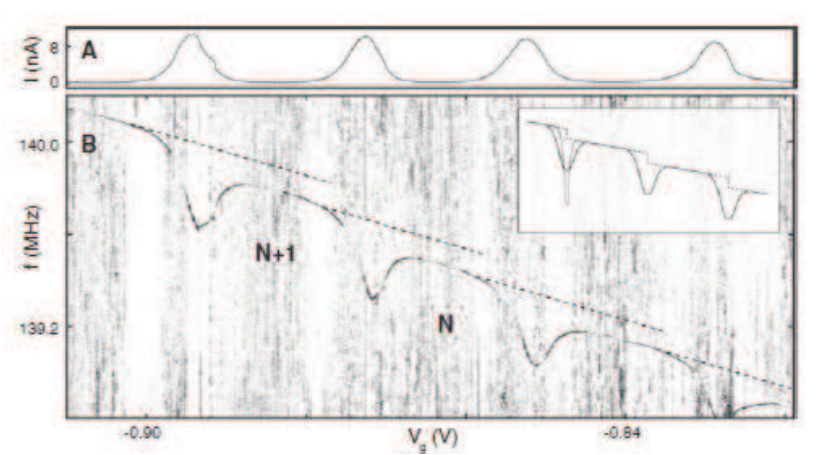} }
\vspace*{-0.cm} \caption{Single electron tuning. (A) Nanotube
current versus gate voltage showing single-electron tunneling at the
peaks and Coulomb blockade in the valleys. (B) Normalized resonance
signal versus rf frequency and gate voltage ($V_{sd}=1.5$~mV). The
tuned mechanical resonance shows up as the darker curve with dips at
the Coulomb peaks. Reprinted with permission from \cite{der},
G.~A.~Steele {\em et al.},  {Science} {\bf 325}, 1103 (2009).
$\copyright$ 2009, American Association for the Advancement of
Science.}
\end{figure}

One of the most promising objects to be used for observing electron shuttling
is a suspended carbon nanotube. To observe an electron shuttle it is
important to fabricate an electromechanical system with two properties:
(i) a soft mode of mechanical vibrations that is susceptible to becoming
unstable when a bias voltage is applied, and (ii) a sufficiently strong
electromechanical coupling that can overcome mechanical damping.
In devices based on suspended carbon nanotubes both these conditions
are realized  \cite{sapmaz, der}. Three different vibrational modes
(the radial mode $\omega_0=const$, the stretching mode
$\omega_0\propto1/L$ and the bending mode $\omega_0\propto1/L^2$,
$L$ is the nanotube length) can be excited in transport experiments
with suspended carbon nanotubes. Recently it was shown that in
a suspended carbon nanotube actuated by rf-radiation, the mechanical
subsystem is sensitive to single-electron transport \cite{der} (see
Fig.~26). In Ref.~\onlinecite{der} the influence of a gate voltage on the
frequency of vibrational modes in the Coulomb blockade regime of
electron transport was studied. The bending mode was excited by
radiation of a nearby rf antenna at the resonance frequency. This
frequency was shown to drop at bias voltages such that the Coulomb blockade
is lifted and an additional electron is transferred through the
nanotube. The experiment demonstrates strong electromechanical
coupling in suspended carbon nanotubes.

\section{Conclusions}

 This review demonstrates that during the last several years there has been
 significant activity in the study of nanoelectromechanical (NEM) shuttle structures.
 New physics was harvested for theoretical suggestions of how to achieve new
 functionality of shuttle devices. A short list of these suggestions would include
 the following items:

1) The polaronic approach for studying nanoelectromechanics beyond
the limit of weak NEM coupling was developed. This approach,
resulting in a qualitative modification of the voltage and
temperature control on NEM performance, could be applied to
molecular-based shuttle devices known for their high mechanical
deformability. 2) New NEM phenomena are possible in superconducting
shuttle devices. The superconducting current through Josephson weak
links supported by Andreev bound states \cite{andreev} (see also
\cite{blom}, \cite{grinzwajg}) is qualitatively modified by coupling
the latter to nanomechanical vibrations. Both mechanically assisted
Cooper-pair transfer and electronically assisted cooling of a
nanomechanical subsystem become possible. Strong electron-vibron
interactions are manifested in the Franck-Condon blockade of the
critical current ($T=0$) and its nonmonotonous (anomalous)
temperature behavior.  3) The possibility to achieve an external
control on the direction of energy transfer between electronic and
mechanical subsystems has been demonstrated. Efficient ground state
cooling of a nanomechanical resonator was predicted in both
superconducting and non-superconducting NEM-SET systems
\cite{cooling, coolsantandrea}. 4) The role of the electronic spin
is important in shuttle devices made of magnetic materials.
Spin-dependent exchange forces can be responsible for a
qualitatively new nanomechanical performance opening a new field of
study that can be called spintro-mechanics. 5) Electronic many-body
effects, appearing beyond the weak tunneling approach, result in
single electron shuttling assisted by Kondo-resonance electronic
states. The possibility to achieve a high
sensitivity to coordinate displacement in electromechanical
transduction along with the possibility to study the kinetics of the
formation of many-body Kondo states has also been demonstrated
recently.

There are still a number of unexplored shuttling regimes  and
systems, which one could focus on in the nearest future. In
addition to magnetic and superconducting shuttle devices one could
explore hybrid structures where the source/drain and gate electrodes
are hybrids of magnetic and superconducting materials. Then one
could expect spintromechanical actions of a supercurrent flow as
well as superconducting proximity effects in the spin dynamics in
magnetic NEM devices. An additional direction is the study of
shuttle operation under microwave radiation. In this respect
microwave assisted spintromechanics is of special interest due to
the possibility of microwave radiation to resonantly flip electronic
spins. As in ballistic point contacts \cite{grinzwajg} such flips
can be confined to particular locations by the choice of microwave
frequency, allowing for external tuning of the spintromechanical
dynamics of the shuttle. Polaronic effects in superconducting and
magnetic shuttle devices represents another interesting direction
for the future research. 
Promising nano-objects where charge
shuttling could play a significant role {\em include} 
bio-systems. 
{\em Electromechanical} phenomena {\em involving both electrons and protons} could be very important for
"transport" effects in cells and bacteria (see
\cite{bio1,bio2,bio3}).

Experimental studies of shuttle devices have been developed
significantly during the last couple of years. A number of new implementations of
the idea of nanomechanical shuttling has been suggested (see chapter V).
The aim of having fully controllable mechanics of the shuttle motion in
combination with the lowest possible level of dissipation motivates the
modifications which have been made. The possibility to achieve
mechanically controllable shuttling of a single electron was a
challenge for a number of years. Evidence indicating that a Coulomb blockade
shuttle device has been made was reported recently \cite{kim}.
Although the regime of self supported shuttle vibrations has been
reached in a number of experiments (see Section 5), the observation of a shuttle
instability still remains a challenge for the future. The
achievements made in the technology of producing shuttle devices
\cite{self} give promises for the construction of magnetic and
superconducting shuttle structures, which would enable one to
explore a number of tempting functionalities theoretically suggested
during recent years.

\section{Acknowledgements} \label{c}

Financial support from the Swedish VR, and the Korean WCU program
funded by MEST/NFR (R31-2008-000-10057-0) is gratefully
acknowledged. This research was supported in part by the Project of
Knowledge Innovation Progra (PKIP) of Chinese Academy of Sciences,
Grant No. KJCX2.YW.W10. I. V. K. and A. V. P. acknowledge financial
support from the National Academy of Science of Ukraine (grant No.
4/12-N). I. V. K. thanks the Department of Physics at the University
of Gothenburg for hospitality. MK is grateful to KITP Santa Barbara
for hospitality. This research was supported in part by the National
Science Foundation under Grant No. NSF PHY11-25915.


\begin{thebibliography}{99}
\bibitem{footnote}
{ Note that in spite of the degeneracy of these two charge states
near either lead, the solution of the Schr{\" o}dinger equation with
the Hamiltonian (\ref{CooppairHam}) corresponds to significant
fluctuations in the number of Cooper pairs on the island.
}
\end{thebibliography}
\end{document}